\documentclass[11pt]{article}
\pdfoutput=1
\usepackage{graphicx,epsfig}
\usepackage{color}
\usepackage[colorlinks,citecolor=red,linkcolor=blue]{hyperref}
\usepackage{slashed}

\usepackage{epsfig,float,amsmath,amsfonts,amssymb,graphicx,bm,bbm,array,dcolumn,mathabx}
\usepackage{subfigure}

\usepackage[T1]{fontenc}
\usepackage[latin9]{inputenc}
\usepackage{varwidth}

\makeatletter

\providecommand{\tabularnewline}{\\}
\newenvironment{cellvarwidth}[1][t]
    {\begin{varwidth}[#1]{\linewidth}}
    {\@finalstrut\@arstrutbox\end{varwidth}}

\usepackage{listings}

\usepackage{ae,aecompl}
\usepackage{multirow}

\usepackage{bbm}

\makeatother

\usepackage{babel}

\usepackage[left=2.5cm, right=2.5cm, top=2cm]{geometry}
\usepackage{amsmath, amssymb, mathrsfs, graphics, setspace}
\newcommand{\mathsym}[1]{{}}
\newcommand{\unicode}[1]{{}}

\def\beq{\begin{equation}}
\def\eeq{\end{equation}}
\def\bea{\begin{eqnarray}}
\def\eea{\end{eqnarray}}
\def\bmat{\begin{pmatrix}}
\def\emat{\end{pmatrix}}
\newcommand{ \slashchar }[1]{\setbox0=\hbox{$#1$}   
   \dimen0=\wd0                                     
   \setbox1=\hbox{/} \dimen1=\wd1                   
   \ifdim\dimen0>\dimen1                            
      \rlap{\hbox to \dimen0{\hfil/\hfil}}          
      #1                                            
   \else                                            
      \rlap{\hbox to \dimen1{\hfil$#1$\hfil}}       
      /                                             
   \fi}                                             %

\def\to{\rightarrow}
\def\Chi{{\cal X}}
\def\Dp{{D}}
\def\Up{{U}}

\def\LIntVV{{\cal L}^{\rm VV}_{\rm int}}

\def\LDirMass{{\cal L}_{\rm Dirac\, mass}}
\def\LMajMass{{\cal L}_{\rm Maj\, mass}}

\def\rMLtxt{M_\chi/\Lambda}

\def\muX{\mu_{\sssty \chi}}
\def\muU{\mu_{\sssty \Up}}
\def\muUB{\mu_{\sssty \bar{\Up}}}
\def\muD{\mu_{\sssty \Dp}}
\def\muDB{\mu_{\sssty \bar{\Dp}}}

\def\MQh{\hat{M}_{\sssty Q}}
\def\muQ{\mu_{\sssty Q}}
\def\muQB{\mu_{\sssty \bar{Q}}}
\def\muQh{\hat{\mu}_{\sssty Q}}

\def\matel#1#2#3{\left< #2| #1 | #3 \right>}

\def\nnbar{$n\!-\!\bar{n}$}

\def\Hh{\hat{H}}
\def\Gm{\Gamma}

\def\Gmzh{\hat\Gamma_0}

\def\GmTA{\langle\Gamma\rangle}
\def\GmcTA{\langle\Gamma^c\rangle}
\def\GmTAh{\langle\hat\Gamma\rangle}
\def\GmBTA{\langle\bar\Gamma\rangle}
\def\GmzTA{\langle\Gamma_0\rangle}
\def\GmPTA{\langle\Gamma'\rangle}
\def\GmPTAh{\langle\hat\Gamma'\rangle}
\def\GmzPTA{\langle\Gamma'_0\rangle}
\def\GmzTAh{\langle\hat\Gamma_0\rangle}
\def\GmzPTAh{\langle\hat\Gamma'_0\rangle}
\def\GmXTA{\langle\Gamma_{\!\!\Chi}\rangle}

\def\GmEffTAh{\langle\hat\Gamma_{\rm\! eff}\rangle}

\def\GmzEffTAh{\langle\hat\Gamma_{0\,{\rm eff}}\rangle}

\def\GmzPEffTAh{\langle\hat\Gamma'_{0\,{\rm eff}}\rangle}

\def\GmSigTA{\langle\Gamma^{(\sigma)}\rangle}

\def\GmzSigTA{\langle\Gamma_0^{(\sigma)}\rangle}

\def\GmzPSigTA{\langle{\Gamma'_0}^{\!(\sigma)}\rangle}

\def\GmzSigTAh{\langle\hat\Gamma_0^{(\sigma)}\rangle}
\def\GmzPSigTAh{\langle{\hat\Gamma_0}'^{(\sigma)}\rangle}
\def\GmSigBTA{\langle\bar\Gamma^{(\sigma)}\rangle}

\def\GmzSigEffTAh{\langle\hat\Gamma^{(\sigma)}_{0\,{\rm eff}}\rangle}

\def\GmzSigPEffTAh{\langle\hat\Gamma'^{(\sigma)}_{0\,{\rm eff}}\rangle}
\def\DGm{\Delta\Gamma}

\def\DGmTA{\langle\Delta\Gamma\rangle}
\def\DGmzoh{\Delta\hat\Gamma_{01}}
\def\DGmzohh{\Delta\hat{\hat\Gamma}_{01}}

\def\DGmzoTAh{\langle\Delta\hat\Gamma_{01}\rangle}
\def\DGmzoTAhh{\langle\Delta{\hat{\hat\Gamma}}_{01}\rangle}

\def\DGmzoEffTAhh{\langle\Delta\hat{\hat\Gamma}_{01\,{\rm eff}}\rangle}
\def\DGmSigTA{\langle\Delta\Gamma^{(\sigma)}\rangle}

\def\DGmSigzoTAh{\langle\Delta\hat\Gamma_{01}^{(\sigma)}\rangle}

\def\DGmSigzoEffTAhh{\langle\Delta\hat{\hat\Gamma}_{01\,{\rm eff}}^{(\sigma)}\rangle}
\def\SigvTA{\langle\sigma\, v\rangle}

\def\Sigvz{\sigma_0 v}
\def\SigvzTA{\langle\sigma_0 v\rangle}
\def\SigvzTAh{\langle\hat\sigma_0 v\rangle}

\def\DSigv{\Delta\sigma\, v}
\def\DSigvTA{\langle\Delta\sigma\, v\rangle}

\def\DSigvzoTAh{\langle\Delta\hat\sigma_{01} v\rangle}
\def\DSigvzoTAhh{\langle\Delta\hat{\hat\sigma}_{01} v\rangle}

\def\eQ{e_Q}
\def\eQB{\bar{e}_{\sssty Q}}
\def\nX{n_{\sssty \Chi}}
\def\nXz{n_{\sssty \Chi}^{\sssty (0)}}
\def\nXeq{n_{\sssty \Chi}^{\sssty\rm (eq)}}
\def\nU{n_{\sssty \Up}}
\def\nD{n_{\sssty \Dp}}
\def\nUB{n_{\sssty \bar\Up}}
\def\nDB{n_{\sssty \bar\Dp}}
\def\nDU{n_{\sssty \Delta\Up}}
\def\nDD{n_{\sssty \Delta\Dp}}
\def\nB{n_{\sssty B}}
\def\nQ{n_{\sssty Q}}
\def\nQB{n_{\sssty \bar Q}}
\def\nDQ{n_{\sssty \Delta Q}}
\def\nQz{n_{\sssty Q}^{\sssty (0)}}
\def\nQeq{n_{\sssty Q}^{\sssty\rm (eq)}}
\def\nQBeq{n_{\sssty \bar{Q}}^{\sssty\rm (eq)}}
\def\nBpL{n_{\sssty B+L}}
\def\nBmL{n_{\sssty B-L}}
\def\nBe{{n_{\sssty B}}_{\sssty e}}
\def\nLe{{n_{\sssty L}}_{\sssty e}}
\def\nBs{{n_{\sssty B}}_{\sssty s}}
\def\nLs{{n_{\sssty L}}_{\sssty s}}
\def\nps{{n_p}_s}
\def\nns{{n_n}_s}
\def\nes{{n_e}_s}

\def\nDnus{{n_{\sssty \Delta\nu}}_s}
\def\rnp{r_{np}}

\def\lrarr{\leftrightarrow}

\def\sssty{\scriptscriptstyle}

\def\GmzPTAh{\langle\hat\Gamma'_0\rangle}

\def\Ychi{Y_\chi}
\def\YB{Y_{\sssty B}}
\def\YBobs{Y_{\sssty B}^{\rm obs}}
\def\etaB{\eta_{\sssty B}}
\def\etaBobs{\eta_{\sssty B}^{\rm obs}}

\def\Ychiz{Y_\chi^{\sssty (0)}}
\def\YQz{Y_{\sssty Q}^{\sssty (0)}}
\def\YSMz{Y_{SM}^{\sssty (0)}}

\def\seff{s_{\rm eff}}
\def\LamQCD{\Lambda_{\rm QCD}}
\def\nnBmm{{n\bar{n}}}
\def\QhME{\langle \hat{Q}\rangle}
\def\QVV{Q_{\rm\sssty VV}}

\def\suU{s^{\sssty uU}}
\def\sdD{s^{\sssty dD}}

\begin{document}


\title{\normalsize{\bf Baryogenesis from a Majorana Fermion Coupled to Quarks}}

\author{\normalsize{Shrihari Gopalakrishna$^{a,b}$\thanks{shri@imsc.res.in}\ , Rakesh Tibrewala$^c$\thanks{rtibs@lnmiit.ac.in}}
\\
$^a$~\small{Institute of Mathematical Sciences (IMSc), Chennai 600113, India.}\\
$^b$~\small{Homi Bhabha National Institute (HBNI), Anushaktinagar, Mumbai 400094, India.}\\
$^c$~\small{The LNM Institute of Information Technology (LNMIIT), Jaipur 302031, India.}
}

\maketitle

\begin{abstract}

  In the theory with a Majorana fermion ($\Chi$) coupled to quark-like fermions ($Q$)
  via a dimension-six four-fermion vector-vector interaction,  
  we have computed in an earlier work the baryon asymmetry generated
  in the decay and scattering processes of the $\Chi$ with $Q$.
  %
  In this work we consider such processes in the expanding early Universe, 
  set up the Boltzmann equations governing the $\Chi$ and net baryon number densities,    
  and numerically solve them in example benchmark points,
  taking the thermally averaged decay and scattering rates and their temperature dependence from the earlier study.  
  We find that starting from a baryon symmetric Universe at early time,
  the presently observed baryon asymmetry of the Universe (BAU)
  can be explained in this theory over a wide range of mass scales,
  $M_\chi\in (10^4,10^{16})$~GeV for appropriately chosen couplings.
  We find that scattering processes play a crucial role in generating the baryon asymmetry in this theory.  
  We present our results in a general manner that should be useful not just in our theory,
  but also in other related theories that share the essential ingredients. 
  Our results should help guide promising ways to probe such new physics in terrestrial experiments.
  For instance, in regions of parameter space that yield the observed BAU,
  we present the rate for neutron-antineutron (\nnbar) oscillation
  and discuss the prospects for observing this in upcoming experiments. 
    
\end{abstract}




\section{Introduction}
\label{Intro.SEC}

Observationally, it is apparent that we live in a Universe in which there is more matter than antimatter at the present epoch, 
the so-called baryon asymmetry of the Universe (BAU).
If the Big-Bang produced matter and antimatter in equal amounts, from a particle physics perspective,
how at late time the BAU emerges remains an open question.
For the asymmetry to develop, the three Sakharov conditions~\cite{Sakharov:1967dj} must be satisfied in any candidate theory,
namely, $C$ and $CP$ violation, baryon number $B$ violation, and departure from thermal equilibrium. 
Although the standard model (SM) of particle physics has these necessary ingredients for producing an asymmetry,
it is commonly believed that it falls short in explaining the observed BAU by many orders of magnitude. 
Various theories beyond the standard model (BSM) have been put forth to explain the observed BAU~(for reviews,
see for example Refs.~\cite{Kolb:1990vq,Cline:2006ts}). 

In Ref.~\cite{Gopalakrishna:2022hwk} we put forward one such proposal containing a Majorana fermion pair
with indefinite baryon number coupled to quark-like fermions, and showed that the three Sakharov conditions  
can be satisfied in the theory, enabling the generation of the BAU.
In Ref.~\cite{Gopalakrishna:2023mul} we identify processes
with tree and loop contributions
for which the decay or scattering rate is different for the process and conjugate process carrying opposite $B$,
and compute the rate difference. 

Our study here is focused on setting up and solving the Boltzmann equation
for the $\Chi$ and baryon number densities in the high-temperature expanding Universe, 
with the above rates appearing as inputs in the collision terms. 
Starting in a baryon symmetric state and evolving through the epoch when the $\Chi$ departs from thermal equilibrium,
we compute the asymptotic value of the baryon asymmetry,
and identify viable parameter regions that result in the observed BAU.
Although we are motivated by the above theory, we present our results in a manner that is applicable to other related theories.
We study the VV interaction, while in the literature the scalar-scalar (SS) interaction is studied;
for a more comprehensive list of references related to our work, see
Refs.~\cite{Gopalakrishna:2022hwk,Gopalakrishna:2023mul}. 
Furthermore, baryon asymmetry generation in decays is usually studied, 
while we include both decay and scattering processes and consider baryon asymmetry generation directly into quarks.
Ref.~\cite{Baldes:2014rda} includes the multiple operator scattering contribution,
but with the SS interaction. 

The paper is organized as follows: 
In Sec.~\ref{BGChiTh.SEC} we briefly review the theoretical framework that our work is based on.  
In Sec.~\ref{BAUgen.SEC} we describe the cosmological setting and
develop the Boltzmann equations governing the $\Chi$ and baryon number densities in the early Universe. 
In Appendix~\ref{ThDUnivRev.SEC} we review thermodynamics applicable in the early Universe, 
and,
in Appendix~\ref{chiOutOfTEUV.SEC} we specify the way we handle the case when the $\Chi$ is not thermalized at any epoch.
In Sec.~\ref{BEnum.SEC} we present our numerical results,
comparing the decay and scattering rates and the Hubble expansion rate in Sec.~\ref{GmGmSigTA.SEC}, 
showing the solutions of the Boltzmann equations and the BAU we obtain in our theory in Sec.~\ref{BENumSoln.SEC},
and in Sec.~\ref{nnbarOsc.SEC} show the rates for neutron-antineutron oscillation in viable regions of parameter space.
%
In Appendix~\ref{GmSigvxDep.SEC} we give a compilation from Ref.~\cite{Gopalakrishna:2023mul} of 
the decay and scattering rates and the functional form for their temperature dependence that are needed above
as the collision terms in the Boltzmann equations. 
In Sec.~\ref{nBFinWO.SEC} we briefly discuss the possibility of a partial wash-out of the generated baryon number 
from other baryon number violating processes such as the sphalerons. 
In Sec.~\ref{Concl.SEC} we offer our conclusions.


\section{The Theory for Baryon Asymmetry Generation}
\label{BGChiTh.SEC}

In this section we review the theoretical framework of Refs.~\cite{Gopalakrishna:2022hwk,Gopalakrishna:2023mul}
that our work is based on. 
The theory extends the SM with a new electromagnetic (EM) charge neutral Dirac fermion $\chi$ which
couples to an up-type color-triplet quark-like fermion $\Up$ with EM charge $+2/3$, 
and to two color-triplet down-type quark-like fermions $\Dp$ with EM charge $-1/3$,
via a dimension-six four-fermion vector-vector (VV) effective operator
\beq
\LIntVV = \frac{1}{2\Lambda^2}\, [\overline{\Dp^c_b} \, \gamma^\mu \Dp_a ] \ [ \bar{\chi} \, \gamma_\mu \left( g_L P_L + g_R P_R \right) \Up_c ]  \, \epsilon^{abc} + H.c. \ ,
\label{LIntVV.EQ}
\eeq
where
$\Lambda$ is the cut-off scale, 
the subscripts $a,b,c$ on the fields are the color indices, and $D^c$ is the charge conjugate field.
With the usual SM quark-like baryon number ($B$) assignments of $B(\Up)\! =\! B(\Dp)\! =\! +1/3$,
from this interaction we assign $B(\chi)\!=\!+1$. 

The masses for the vector-like fermions $\Up,\Dp,\chi$ are given by  
\bea
    \LDirMass = - M_{\Dp} \overline{\Dp} \Dp - M_{\Up} \overline{\Up} \Up - M_\chi \overline{\chi} \chi \ , \\
    \LMajMass = - \frac{1}{2} \overline{\chi^c} (\tilde{M}_L P_L + \tilde{M}_R P_R)\, \chi + h.c. \ ,
\label{LMassEff.EQ}    
\eea
where $M_U,M_D,M_\chi$ are Dirac masses. 
We also include Majorana masses $\tilde{M}_{L,R}$ for the $\chi$ that break baryon number,
which are the only sources of baryon number violation in our theory, 
and therefore all baryon number rate asymmetries must be proportional to either of these. 
The $\tilde{M}_{L,R}$ splits the Dirac fermion $\chi$ into two Majorana fermions $\Chi_n$, with indefinite baryon number.
We diagonalize the $\chi$ mass matrix and work in the mass basis in which the VV interaction is~\cite{Gopalakrishna:2022hwk} 
\beq
\LIntVV = \frac{\epsilon^{abc}}{2 \Lambda^2} 
\left[  \widebar{\Dp^c_b} \, \gamma^\mu \Dp_a \right] \ \left[ \bar{\Chi}_n \, G^n_{V \mu} \Up_c \right]
+ h.c. 
 \ ,
\label{LIntVVMB.EQ}
\eeq
where the Lorentz structure of the coupling $G_V$ is detailed in Ref.~\cite{Gopalakrishna:2022hwk},
and is written in terms of $g_{\sssty L,R}$ and the elements of the unitary matrix ${\cal U}$ that diagonalizes the mass matrix, 
with combination being denoted as $\hat{g}_{\sssty L,R}$.
The physical phases in the coupling and ${\cal U}$ lead to $C$ and $CP$ violation. 

In Ref.~\cite{Gopalakrishna:2022hwk} we give many examples for how our effective theory could 
be completed into a renormalizable UV-complete theory. 
We argued there that from this perspective, 
in addition to the effective interaction term in Eq.~(\ref{LIntVVMB.EQ}), 
other effective operators could also be generated, namely, 
\beq
    {\cal L} \supset - \frac{1}{\Lambda^2} \, [\bar{\Up}_c \bar{G}^n_{V \mu} \Chi_n]\, [\overline{\Chi}_m {G^m_{V}}^\mu \Up_c] 
    - \frac{1}{2 \Lambda^2} (\delta^{aa'} \delta^{bb'}
    - \delta^{ab'} \delta^{ba'}) [\overline{\Dp}^c_b\, \tilde{g} \gamma^\mu \Dp_a]\, [\bar{\Dp}_{a'}\, \tilde{g} \gamma_\mu \Dp^c_{b'}] \ . 
\label{modAops.EQ}    
\eeq
In this work, we also include these operators.  

The EM charges of the $\Up,\Dp$ allow couplings with the SM $u,d$ quarks.
For instance, in the simple case of the $\Up,\Dp$ being $SU(2)_L$ singlets, we can write interaction terms~\cite{Gopalakrishna:2022hwk}
\beq
    {\cal L}_{\rm Yuk} \supset -\tilde{y}_d^i\, \bar\Dp H^\dagger q_L^i + \tilde{y}_u^i\, \bar\Up H\! \cdot\! q_L^i + {\rm h.c.} \ ,
    \label{LuUdDmix.EQ}
\eeq    
where $H$ is the SM Higgs doublet,
the $q_L^i$ are the three generation SM quark doublets with $i=\{1,2,3\}$ being the SM generation index.
Electroweak symmetry breaking due to $\langle H \rangle = (0\ v/\sqrt{2})^T$ generates mass mixing between
$\Up \lrarr u^i$ and $\Dp \lrarr d^i$,
and combining this with the $u,d$ SM mass terms, we can write in terms of
$(u^i\ \Up)^T$ and $(d^i\ \Dp)^T$ the $4\!\times\! 4$ up and down type mass matrices respectively.
These can be reduced to a block diagonal form by unitary rotations with (the sine of the) mixing angles
$\suU_i$ and $\sdD_i$ respectively (see, for example, Refs.~\cite{Gopalakrishna:2011ef,Gopalakrishna:2013hua} for details),
with the upper $3\!\times\! 3$ block identified with the three generation SM quark $u^i,d^i$ sectors.
The diagonalization of the SM quark sector then proceeds as in the SM to obtain the Cabibbo-Kobayashi-Maskawa (CKM) matrix as usual.
We can infer that the VV interaction of Eq.~(\ref{LIntVVMB.EQ}), due to this mixing, induces the operator with SM quarks
\beq
{\cal L} \supset \frac{\epsilon^{abc}}{2 \Lambda^2}\,\hat{g}_{\sssty L,R} \seff^{ijk}\,
\left[  \widebar{d^c_b}^i \, \gamma^\mu d^j_a \right]\, \left[ \bar{\Chi} \, \gamma_\mu P_{\sssty L,R} u^k_c \right] \ ,
\label{LChiSMudd.EQ}
\eeq
where 
the $i,j,k$ are SM generation indices,
and,
$\seff^{ijk} \equiv (\sdD)^2 \suU V_{ijk}$, the $V_{ijk}$ being combinations of the CKM rotation matrices.

Our focus in this work is on how the $\Chi$ dynamics generates net baryon number in the early Universe.  
The $u^i\lrarr \Up$ and $d^i\lrarr\Dp$ mixings above are baryon number conserving
and only distribute the net baryon number across the SM flavors.
However, the CKM phase may have an effect on the amount of CP violation in the processes that are at play. 
For a study on this, albeit in a different theoretical setting, see for example, Ref.~\cite{Aitken:2017wie}.
When the $\Chi$ is integrated out, the operator in Eq.~(\ref{LChiSMudd.EQ}) with $i\!=\!j\!=\!k\!=\!1$
induces neutron-antineutron oscillation. 
Other choices of $i,j,k$ induce flavor changing observables, which can be searched for in experiments.
We discuss these further in Sec.~\ref{nnbarOsc.SEC}.

In Ref.~\cite{Gopalakrishna:2023mul}, we identify loop contributions in $\Chi$ decay and scattering processes
that generate a baryon asymmetry, 
and compute the rate difference between the various processes and their corresponding conjugate processes,
and show their dependence on the couplings, phases, and masses in the theory
in the pseudo-Dirac limit $(M_n-M_\chi)/M_\chi\!\ll\!1$.
Two classes of loop contributions are identified there, namely,
the single-operator (with only the operator in Eq.~(\ref{LIntVVMB.EQ}))
and the multiple-operator contributions (with the operators in Eqs.~(\ref{LIntVVMB.EQ})~and~(\ref{modAops.EQ})),
that yield a rate difference between process and conjugate process, 
which is computed for sample parameter space points -- benchmark points BP-A and BP-B respectively.  
In addition to these, in this study, we also include another benchmark point, BP-C, as explained in Sec.~\ref{BEnum.SEC}. 

The $\Chi$ decay and scattering processes that generate the baryon asymmetry are set in the high-temperature early Universe.
The $\Chi$ starts out being in thermal equilibrium (TE) and eventually decouples as the temperature falls,
and we track its number density using the Boltzmann equation (cf. Sec.~\ref{BEsetup.SEC}). 
For the relevant decay and scattering processes,
in Ref.~\cite{Gopalakrishna:2023mul}
we compute the decay width and scattering cross section as a function of initial state momenta
and average them over the thermal distribution functions to obtain the thermally averaged rates. 
We summarize those findings in Sec.~\ref{GmGmSigTA.SEC} and Appendix~\ref{GmSigvxDep.SEC}. 

For our study here,
we import from Ref.~\cite{Gopalakrishna:2023mul} the thermally averaged decay and scattering rates, 
keeping their temperature dependence, but making the following simplifying assumptions.  
It is seen there that the rate difference is largely generated in the decay and scattering of the $\Chi_2$,
and we therefore keep only this state in our analysis here, denoting it simply as $\Chi$ with mass $M$.
Furthermore, we do not wish to repeat here how the rates are affected by the coupling structure and phases in $G_V$
as it has already been elucidated in that study, and therefore denote its coupling simply as $g$, 
but retain how the decay and scattering rates scale with $g$ and $\Lambda$. 
Keeping track of the distinction between the $\Up,\Dp$ is also not important for our analysis here,
and we denote them collectively as $Q$ with mass $M_Q$. 
Thus, we abstract from Ref.~\cite{Gopalakrishna:2023mul} the thermally averaged rates, 
$\GmTA$, $\GmSigTA$ and $\GmPTA$, 
for the decay process $\Chi\to QQQ$, the scattering process $\Chi Q^c\to QQ$,
and the $\Delta B\!=\! 2$ scattering process $QQQ \to Q^cQ^cQ^c$, respectively, 
along with the rates for their conjugate and inverse processes, with $Q\!=\!\{\Up,\Dp\}$ appropriately chosen. 
In this work, we consider these processes in the early Universe, 
and compute the BAU from the knowledge of these thermally averaged decay and scattering rates.
We turn next to a discussion of how the BAU comes about in our theory.

\section{The Cosmological Setting}
\label{BAUgen.SEC}

In this section we place the microscopic processes reviewed in Sec.~\ref{BGChiTh.SEC} in a cosmological setting,
and develop the Boltzmann equation (BE) for the net baryon number in the early Universe to determine the BAU.

The BAU today can be expressed as the ratio
\beq
\etaB \equiv \frac{\nB}{n_\gamma} = \frac{(n_b - n_{\bar{b}})}{n_\gamma} \ , {\rm\ or,} \quad \YB\!\equiv\!\frac{(n_b-n_{\bar b})}{s} \, \ ,
\label{etaBDefn.EQ}
\eeq
with $n_b$ ($n_{\bar{b}}$) the number density of baryons (anti-baryons), $\nB$ the net baryon number density, $n_\gamma$ is the number density of photons today, $s$ is the entropy density,
and $\YB$ is the baryon number {\em yield} (see Appendix~\ref{ThDUnivRev.SEC} for a review of thermodynamics applicable in the expanding early Universe).
We can write $\etaB\! \approx\! 6\times 10^{-10}\, \times (\Omega_b h^2/0.022)$ and taking the CMB observations~\cite{Planck:2018vyg} of $\Omega_b h^2\! \approx \!0.022$
we have $\etaBobs \! \approx \! 6\times 10^{-10}$,
which agrees well with the predictions from BBN~\cite{ParticleDataGroup:2024cfk}. 
The observed BAU can equivalently be written as $\YBobs \approx 0.85\times 10^{-10}$. 

In Ref.~\cite{Gopalakrishna:2023mul} we consider the $\Up,\Dp,\Chi$ interacting in the early Universe at temperature $T$, 
and discuss relations between the chemical potentials
enforced by the processes in TE, namely, 
$\muUB\!=\!-\muU$, $\muDB\!=\!-\muD$ due to the QED processes
such as $\Up\bar{\Up}, \Dp\bar{\Dp} \lrarr \gamma, \gamma\gamma$, with $\mu_\gamma \!=\! 0$ as it is a massless real boson,  
and, when the $\Chi$ is in TE, we have $\muX \!=\! 0$ as it is a Majorana fermion. 
As explained there,
with $\nDU \!=\! \nU-\nUB$, $\nDD \!=\! \nD-\nDB$, 
the form of the VV interaction with antisymmetry in color
implies $\nB \!=\! (\nDU+2\nDD)/3$. 
There, we present the BEs for the evolution of the number densities of the $\Chi,\Up,\Dp$
in the expanding Universe as a function of temperature $T$,  
and the BE for $\nB$ in terms of the BEs for $\nU,\nD$. 

In this work we drop the distinction between the $Q=\{\Up,\Dp\}$,
and denote by $Q$ {\em either} of the $\Up$ or $\Dp$, 
which is a good approximation in the $(M_\Up-M_\Dp)/M_\chi\! \ll\! 1$ limit.
In this limit we have $\muU \!\approx\! \muD \equiv \muQ$,
and we have $\nB \!=\! \nDQ/3$, 
with 
$\nQ \!=\! \nU + 2 \nD$, $\nQB \!=\! \nUB + 2 \nDB$, and, $\nDQ \!=\! \nQ-\nQB$. 
The equilibrium number densities are $\nQ \!=\! \nQeq \!=\! \nQz e^{\muQ/T}$ and $\nQB \!=\! \nQBeq \!=\! \nQz e^{\muQB/T}$, 
where $\nQz$ is the number density with zero chemical potential.
Due to the QED processes in equilibrium, we have $\muQB\! =\! -\muQ$. 
We write here the BE for $\nX, \nQ, \nQB$,
and obtain the BE for $\nB$ by taking the difference between the BEs for $\nQ$ and $\nQB$.
The $\Chi$, however, being EM neutral and coupled via the VV interaction,
could behave quite differently depending on how the interaction rate
compares to the Hubble expansion rate.
If the $\Chi$ were to be in TE, the Majorana nature of $\Chi$ implies $\mu_\chi\!=\! 0$
and its number density would be $\nXz$,
and when it departs from TE its number density $\nX$ deviates from $\nXz$. 

We start in a baryon-antibaryon symmetric early Universe, i.e. $\YB(T\!\gg\! M_\chi)\! =\! 0$,
at a temperature much above the electroweak symmetry breaking scale of the order of a TeV. 
%
%
As $T$ falls in the expanding Universe,
it is the $\Chi$ that has a chance to deviate from TE first, i.e. to decouple from the thermal plasma,
due to its weaker effective coupling,  
when the $\Chi$ decay and scattering rates and inverse rates become smaller than the Hubble expansion rate $H(T)$ (cf. Sec.~\ref{GmGmSigTA.SEC}). 
We expect the $\Chi$ to eventually decouple for some $x \gtrsim 1$, where $x\equiv M_\chi/T$, 
since the interaction rates
of the $\Chi$ with $Q$ becomes smaller than $H$ for some large enough $x$,
due to its dependence on $\nQ$ (cf. Sec.~\ref{BEsetup.SEC})
with $\nQ = \nQ^{(eq)} \sim e^{-x}$ being exponentially suppressed.\footnote{The $K$ defined in Ref.~\cite{Kolb:1990vq}, in our notation, is
$K\! \equiv\! \Gamma/H(M_\chi) \! =\! (1/2)\GmEffTAh_i(x\!=\!1)$ (cf. p.\,\pageref{GmEffDefn.PG}).
In Ref.~\cite{Kolb:1990vq} the generation of the baryon asymmetry in decays for various $K$ are shown. 
The leptogenesis framework in Ref.~\cite{Fukugita:1986hr} is in the $K < 1$ regime.
For an early numerical study on leptogenesis, see Ref.~\cite{Luty:1992un}.  
In our work here, we also have $K < 1$,
but in addition to decay, we include scattering processes also, with the latter playing a dominant role in many of our benchmark points.}
Given the $C$, $CP$, and $B$ violation in our theory,
all three Sakharov conditions are satisfied, and a baryon asymmetry can build up, which we observe at $T \ll M_\chi$ as the BAU.
The baryon asymmetry builds up only when the $B$-violating processes decouple,
as otherwise the forward and backward reactions are in TE and will drive the $B$-number to zero
(in other words, in TE,
the chemical potential corresponding to baryon number $\mu_B\! =\! 0$ in the presence of efficient baryon number violating processes).
%
As discussed in Ref.~\cite{Gopalakrishna:2022hwk} and reviewed in Sec.~\ref{BGChiTh.SEC}, 
the $Q$ mixes with the SM quarks and thus eventually decays to the latter in a baryon number conserving way, 
transferring faithfully the net baryon number generated by the $\Chi$ dynamics to the SM quarks.  

Depending on the parameters, it is possible that $\Chi$ is never in TE in the relevant $x$ range.
In fact, if it is not in TE for $x=1$, we can infer that it is not so for any $x<1$
since at best $\GmSigTA/M_\chi \sim 1/x$ while $H/M_\chi \sim 1/x^2$.
Not having the anchor of being in TE at some high temperature to set $\nX = \nXeq$ 
means that the $\Chi$ abundance becomes dependent on the physics of the UV completion that produced the $\Chi$.
In this case, no prediction is possible in a model independent way,
and whether the correct $\YB$ is produced has to be directly answered in each specific UV completion.
However, if we assume that the $\Chi$ was produced by some other state in the UV completion
that was in TE, some predictivity can be salvaged.
We discuss an example of this kind in a UV completion example in Appendix~\ref{chiOutOfTEUV.SEC}, 
which we give as a proof of principle. 
To be more general, when the $\Chi$ is not in TE, we treat $\delta_\Chi(x_b)\!\equiv\!\nX(x_b)/\nXz(x_b)$ as a free parameter. 
$\delta_\Chi(x_b) \neq 1$ then encodes that the $\Chi$ is not in TE, and by how much the $\nX(x_b)$ deviates from $\nXz(x_b)$.
For various choices of the initial condition $\delta_\Chi(x_b)$, we solve the Boltzmann equations
and compute the late-time baryon asymmetry that is generated in our theory.

We turn next to giving details of the Boltzmann equations in the expanding Universe. 

\subsection{The Boltzmann Equations}
\label{BEsetup.SEC}

Here we derive the Boltzmann Equation (BE) for the $\Chi$ number density $n_\chi$,
and for the net baryon number density $\nB$ in the (expanding) early Universe
(for a review, see for example, Ref.~\cite{Kolb:1990vq}, and Appendix~\ref{ThDUnivRev.SEC}).
The amount of the BAU generated in our theory depends on the parameters of our theory reviewed in Sec~\ref{BGChiTh.SEC}, namely,
the mass scale $M_\chi$, the cutoff scale $\Lambda$, and, the coupling $g$. 

We list next the relevant reaction rates that are the collision terms on the rhs of the BE.
For the decay process $\Chi \to QQQ$ and its conjugate process $\Chi \to Q^cQ^cQ^c$
with the zero-$T$ rest-frame decay widths $\Gm$ and $\Gm^c$ and the decay width difference $\DGm = \Gm-\Gm^c$,  
we obtain the thermally averaged decay rate and decay rate difference as~\cite{Gopalakrishna:2023mul}
\beq
\GmTA = f_{TD} \Gm \ ; \ \GmcTA = f_{TD} \Gm^c \ , \ \DGmTA = f_{TD} \Delta\Gamma \ , \
\ {\rm where}\ f_{TD}(x) \equiv \frac{M}{\left<E(p)\right>} =  \frac{K_1(\hat{M} x)}{K_2(\hat{M} x)} \ , 
\label{CGmTADefn.EQ}
\eeq
where
$x=M_\chi/T$, $\hat{M} = M/M_\chi$,
and,
$K_{1,2}$ are the modified Bessel functions of the second kind.
$f_{TD}$ is nothing but the thermal average of the (inverse of the) time-dilation factor. 
We use $\bar{Q}$ and $Q^c$, or $\bar\Gm$ and $\Gm^c$ interchangeably as the distinction is unimportant when dealing with spin-summed rates. 
The total decay rate is $\GmXTA = \GmTA + \GmBTA$.
Thus, given the zero-temperature decay rate $\Gm$ and rate difference $\DGm$, for instance from our study in Ref.~\cite{Gopalakrishna:2023mul}, 
we obtain their thermal averages $\GmTA,~\DGmTA$ by simply multiplying by the $f_{TD}(x)$ factor above. 
The inverse-decay channels are
$QQQ \to \Chi$ and $Q^cQ^cQ^c \to \Chi$ respectively. 

\label{procsDef.PG}
The thermally averaged scattering rate is
$\GmSigTA = \left<\sigma_{\Chi\bar{Q}} v\right> \nQz$, given in terms of the scattering cross section, 
and
for the conjugate process is $\GmSigBTA = \left<\sigma_{\Chi Q} v\right>  \nQz$. 
The relevant scattering cross sections are   
$\sigma_{\Chi\bar{Q}} \equiv \sigma(\Chi Q^c \to Q Q)$, 
$\sigma_{\Chi Q} \equiv \sigma(\Chi Q \to Q^c Q^c)$, 
with their inverse scattering channel cross sections being
$\sigma_{Q Q} \equiv \sigma(Q Q \to \Chi Q^c)$,
$\sigma_{\bar{Q}\bar{Q}} \equiv \sigma(Q^c Q^c \to \Chi Q)$ respectively, 
and the difference in scattering rates is
$\DGmSigTA = \DSigvTA \nQz$ with $\Delta\sigma = \sigma_{\Chi\bar{Q}} - \sigma_{\Chi Q}$.
We denote the thermally averaged LO scattering rate of the $\Delta B\!=\! 2$ process
$QQQ\to Q^cQ^cQ^c$ as $\GmzPTA$, and that of $QQ\to Q^cQ^cQ^cQ^c$ as $\GmzPSigTA$. 
The inverse channel thermal averages are written in terms of $n_\chi^{\sssty (0)}$
by using energy conservation in the statistical distribution functions~\cite{Kolb:1990vq}.

Although our analysis is quite general, our focus will be on the model
reviewed above in Sec.~\ref{BGChiTh.SEC}, taking into account the rates for the
decay process $\Chi \to QQQ$ 
(with final state $B=+1$) and 
$\Chi \to Q^cQ^cQ^c$ 
(with final state $B=-1$), 
the scattering process $\Chi Q^c\! \to\! QQ$ and its conjugate,
and, the $\Delta B\!=\!\pm 2$ scattering process $QQQ\! \to\! Q^cQ^cQ^c$ and its conjugate.
We follow the $\Chi,Q,Q^c$ number densities as $T$ falls, 
and determine the final baryon number yield $\YB(T\!\ll \! M_\chi)$.

The assumption of $CPT$ invariance implies relations among the decay and inverse-decay matrix elements
(mod-squared, summed over spins),
which we explain in detail in Ref.~\cite{Gopalakrishna:2023mul}. 
The relations we obtain are,  
$|{\cal M}(\Chi \to QQQ)|^2 = |{\cal M}(Q^cQ^cQ^c \to \Chi)|^2$,
and
$|{\cal M}(\Chi \to Q^cQ^cQ^c)|^2 = |{\cal M}(QQQ \to \Chi)|^2$, 
and between the inverse and forward scattering channel matrix elements 
$|M|^2_{\bar{Q}\bar{Q}} = |M|^2_{\Chi \bar{Q}}$ and $|M|^2_{Q Q} = |M|^2_{\Chi Q}$,
labeled using the same notation as for the cross sections above. 
After folding in the thermal distribution functions in the initial state and the final state phase space
and integrating, these relations also relate the corresponding thermally averaged rates. 

The BE governing $\nX$, $\nQ$, $\nQB$ are~\cite{Gopalakrishna:2022hwk, Gopalakrishna:2023mul}
\bea
\frac{d}{dt}\nX + 3 H \nX &=& - \GmzTA \left[2\nX - (\eQ^3+\eQB^3) \nXz\right]
                                  - \GmzSigTA \left[(\eQ+\eQB) \nX - (\eQ^2+\eQB^2) \nXz\right] \ ,  \label{BEchi.EQ} \\
\frac{d}{dt} \nQ + 3 H \nQ &=& 3\left[ \GmTA\, \nX - \eQ^3 \nXz \GmBTA
+ \left(\eQB \GmSigTA - \eQ \GmSigBTA \right) \nX + \left(\eQB^2 \GmSigTA - \eQ^2 \GmSigBTA \right) \nXz \nonumber \right. \\ 
&& \left.
- \left(\eQ^3 \GmzPTA\, + \eQ^2 \GmzPSigTA\right) \nXz + \left(\eQB^3 \GmzPTA\, + \eQB^2 \GmzPSigTA\right) \nXz \right]  \ , \label{QBoltzEq.EQ} \\
\frac{d}{dt} \nQB + 3 H \nQB &=& 3\left[ \GmBTA\, \nX - \eQB^3 \nXz \GmTA 
+ \left(\eQ \GmSigBTA - \eQB \GmSigTA \right) \nX  + \left(\eQ^2 \GmSigBTA - \eQB^2 \GmSigTA \right) \nXz \nonumber \right. \\
&& \left.
- \left(\eQB^3 \GmzPTA\, + \eQB^2 \GmzPSigTA\right) \nXz + \left(\eQ^3 \GmzPTA\, + \eQ^2 \GmzPSigTA\right) \nXz \right] \ , \label{QBBoltzEq.EQ}
\eea
where
$\eQ\! =\! e^{\muQ/T}$, $\eQB\! =\! e^{-\muQ/T}$,  
$H$ is the Hubble expansion rate (cf. Eq.~(\ref{HofT.EQ})),
and, 
$\nXz$ is the equilibrium number density (cf. Eq.~(\ref{neqOfT.EQ})). 
In Eq.~(\ref{BEchi.EQ}), the BE for $\nX$,
(but not in the BE for $\nQ,~\nQB$),
we take the rates at leading order (LO), i.e. $\GmTA \approx \GmBTA = \GmzTA$ and $\GmSigTA \approx \GmSigBTA = \GmzSigTA$, 
as it is sufficient to calculate the baryon asymmetry to first order given that the observed BAU is very small, 
and keeping these differences in the $n_\chi$ BE only contributes a second order effect in the baryon asymmetry.

The $\Chi$ is decoupled if the interaction rates are less than $H(T)$, 
i.e. if $\GmTA, \GmSigTA \lesssim H$.
If $\Chi$ is decoupled, $\nX$ could deviate significantly from $\nXz$
and is not exponentially suppressed like $\nXz$ (and $\nQ$) for $M_\chi/T \gtrsim 1$.
If decay and scattering processes involving the $\Chi$ are not in TE,  
the Sakharov condition requiring a departure from TE could be satisfied,
and a nonzero baryon number could develop.

In our theory, the baryon asymmetry generation mechanism in decay and scattering discussed in
Refs.~\cite{Gopalakrishna:2022hwk,Gopalakrishna:2023mul} leads to $\nQ \neq \nQB$. 
The BE for the net baryon number density
$\nB = (\nQ - \nQB)/3$
we obtain from Eqs.~(\ref{QBoltzEq.EQ})~and~(\ref{QBBoltzEq.EQ}) is
\bea
\frac{d}{dt} \nB + 3 H \nB &=& \DGmTA \left( \nX - \nXz \cosh{\frac{3\muQ}{T}} \right)
    + \DGmSigTA \left( \nX \cosh{\frac{\muQ}{T}} - \nXz \cosh{\frac{2\muQ}{T}} \right)
    \nonumber \\ &&
    - 2 \left(\GmzTA + 2\GmzPTA\right)\, \nXz \sinh{\frac{3\muQ}{T}}
    - 2 \GmzSigTA \left(\nX \sinh{\frac{\muQ}{T}} + \nXz \sinh{\frac{2\muQ}{T}} \right)
    \nonumber \\ &&
    -4 \GmzPSigTA\, \nXz\, \sinh{\frac{2\muQ}{T}} \ .
\label{BEnb.EQ}
\eea
We have
$\sinh{(\muQ/T)} = (1/2)\, \nB/\nQz$, 
and the chemical potential corresponding to baryon number $B$ is $\mu_B\! =\! 3 \muQ$.

The sign of the $\nXz$ term in the first term of Eq.~(\ref{BEnb.EQ}) is opposite to that obtained by a naive calculation. 
We discuss in detail in Ref.~\cite{Gopalakrishna:2023mul} 
how CPT and unitarity considerations bring about an intricate relationship between
the inverse decay ($QQQ\to \Chi$) and the $\Delta B\! =\! \pm 2$ scattering channel $QQQ\to Q^cQ^cQ^c$ and its inverse,
and show that the real (on-shell) intermediate state (RIS) contribution in the scattering channel 
overturns the sign of the inverse decay contribution
and leads to the $(\nX - \nXz \cosh{3\muQ/T})$ form shown in Eq.~(\ref{BEnb.EQ}), 
leaving the pure scattering (i.e. without the RIS piece) tree-level $\left<{\Gamma_0'}\right>$ term. 
Ref.~\cite{Kolb:1979qa} explains this in a toy model, 
giving the result to first order in $\muQ/T$, which we generalize to the form shown.
Furthermore, in Ref.~\cite{Gopalakrishna:2023mul},
we extend the result to the inverse scattering channels also taking into account the RIS pieces coming from the
$QQ\to Q^c Q^c Q^c Q^c$ scattering channel,
which overturns the sign of the $QQ\to \Chi Q^c$ inverse scattering channel resulting in the form shown in the BE,
leaving the pure scattering $\GmzPSigTA$ term. 


The leading contribution to the decay rate difference $\DGm$ is given by the interference term between the tree and loop amplitudes
and is given as $\DGm\! =\! -2\DGmzoh$ as shown in Refs.~\cite{Gopalakrishna:2022hwk,Gopalakrishna:2023mul}. 
Taking thermal averages of this leads to the thermally averaged decay rate difference between the process and conjugate process given as $\DGmTA\! =\! -2\DGmzoTAh$.
Similarly, the cross section difference is given in terms of the interference terms of the tree and loop-level scattering amplitudes,
leading to the cross section difference $\DSigvTA = -2\DSigvzoTAh$.
The scattering rate is given as 
$\GmSigTA = (2 \pi^2 g_{*S}/45) M_\chi^3 \SigvTA Y_Q^{(0)}/x^3$,
and similarly for the scattering rate difference
$\DGmSigTA$ with $\DSigvTA$ on the rhs, or $\DGmSigzoTAh$ with $\DSigvzoTAh$ on the rhs, 
where these are related as $\DGmSigTA\! =\! -2 \DGmSigzoTAh$. 
We define
$\GmEffTAh_i(x)\! =\! \GmTA_i(x)/H(M_\chi)\! =\! 1/(1.66 \sqrt{g_*})\, (M_{Pl}/M_\chi) (\GmTA_i/M_\chi)$ for each of these $\GmTA_i$, 
to obtain the corresponding dimensionless rates 
$\left\{\GmzEffTAh, \DGmzoEffTAhh, \GmzSigEffTAh,\right.\allowbreak\left. \DGmSigzoEffTAhh,\GmzPEffTAh, \GmzSigPEffTAh \right\}$ respectively.
\label{GmEffDefn.PG}

We write the BE equivalently in terms of the yield variables $Y_{\chi}\!=\! n_{\chi}/s$ and $Y_{B}\!=\! n_{B}/s$,
where $s$ is the entropy density (cf. Eq.~(\ref{sngamOfT.EQ})).
We also change the independent variable from $t$ to $T$ using Eq.~(\ref{tToTRD.EQ}) 
and then to $x$.
From Eq.~(\ref{YeqOfT.EQ}), we have 
$Y^{(0)}_{\chi,Q} = n^{(0)}_{\chi,Q}/s = 45/(4\pi^4)\, g/g_{*S}\, \hat{M}_{\chi,Q}^2\, x^2 K_2(\hat{M}_{\chi,Q}\, x)$,
where $\hat{M}_Q\!\! =\!\! M_Q/M_\chi$, and $\hat{M}_\chi\!\! =\!\! 1$.
In terms of the above defined quantities, from Eqs.~(\ref{BEchi.EQ}) and (\ref{BEnb.EQ}), 
the BE now is
\bea
\frac{d}{dx} Y_\Chi\!\!\!\! &=&\!\!\!\! -2x\, \left[ \GmzEffTAh (Y_\Chi - \cosh{(3\muQh x)}\, Y_\Chi^{(0)})
                                   +\GmzSigEffTAh (Y_\Chi\cosh{(\muQh x)} - \cosh{(2\muQh x)}\, Y_\Chi^{(0)}) \right] \ , \\
\frac{d}{dx} \YB\!\!\!\! &=&\!\!\!\! -2x \,
\left[
  \DGmzoEffTAhh \, \left(Y_\Chi - \cosh{(3\muQh x)}\, Y_\Chi^{(0)} \right)
 + \DGmSigzoEffTAhh \left(\cosh{(\muQh x)}\, Y_\Chi - \cosh{(2\muQh x)}\, Y_\Chi^{(0)} \right)
 \right. \nonumber \\ && \left.
\hspace*{0.5cm} + \left(\GmzEffTAh + 2 \GmzPEffTAh \right) \, \sinh{(3\muQh x)}\, Y_\Chi^{(0)}  
 + \GmzSigEffTAh \left(\sinh{(\muQh x)}\, Y_\Chi + \sinh{(2\muQh x)}\, Y_\Chi^{(0)} \right)
 \right. \nonumber \\   && \left.  
\hspace*{0.5cm} +2 \sinh{(2\muQh x)} \GmzSigPEffTAh Y_\Chi^{(0)}
 \right]
\ , 
\label{BEYchiYB.EQ}
\eea
where we define the dimensionless $\muQh = \muQ/M_\chi$.  
We have $\sinh(\muQ/T) = \sinh(\muQh x) = (1/2)\, \YB/Y_Q^{(0)}$.\footnote{
We note the identity $\sinh{3x} = 4 \sinh^3{\!x} + 3 \sinh{x} \approx 3 \sinh{x}$, and, $\sinh{2 x} = 2 \sinh{x} \cosh{x} \approx 2 \sinh{x}$.
}
%
Since the observed BAU is tiny, it is sufficient to take the BE to first order in $\YB$, or equivalently in $\muQ$ or $\mu_B$. 
Therefore, we can take $\cosh{(...)} \approx 1$ and $\sinh(\muQh x) \approx \muQh x$ etc., 
and obtain the BE in this approximation as
\bea
\frac{d}{dx} Y_\Chi &\approx& -2x\, \big(\GmzEffTAh + \GmzSigEffTAh)(Y_\Chi - Y_\Chi^{(0)} \big) \ , \\
\frac{d}{dx} \YB &\approx&  -2x \,
\big[
  \big(\DGmzoEffTAhh + \DGmSigzoEffTAhh \big) \, \big(Y_\Chi - Y_\Chi^{(0)} \big)
  \\  
  && 
  + \big(\GmzEffTAh + 2 \GmzPEffTAh \big) \, Y_\Chi^{(0)} \frac{3}{2} \frac{\YB}{Y_Q^{(0)}}
  + \GmzSigEffTAh \big(\frac{1}{2} Y_\Chi + Y_\Chi^{(0)} \big) \frac{\YB}{Y_Q^{(0)}}
  + 2 \GmzSigPEffTAh Y_\Chi^{(0)} \frac{\YB}{Y_Q^{(0)}}
  \big]
    \ . \nonumber
\label{BEYchiYBLO.EQ}
\eea

We find the $QQQ\!\to\! Q^cQ^cQ^c$ scattering rate $\GmzPTAh$ comparable or smaller than the $\GmzTAh$.
Also, we expect the $QQ \to Q^cQ^cQ^cQ^c$ scattering rate $\GmzPSigTAh$ of the same size as $\GmzPTAh$ at best,
if not smaller due to an extra phase space suppression factor, 
and we therefore do not include it in our analysis. 
%

We turn next to solving these coupled set of BE for the late-time $\YB$, 
to find regions of parameter space that yield the observed BAU.

\section{Numerical Results}
\label{BEnum.SEC}

As reviewed in Sec.~\ref{BGChiTh.SEC}, 
in Ref.~\cite{Gopalakrishna:2023mul} we identify processes for which the decay and scattering rate in our new physics theory
is different for the process and conjugate process,
and compute the rate difference for two benchmark points (BPs), namely, 
BP-A for the two-loop single operator contribution, and BP-B for the one-loop multiple operator contribution.
Here we add another benchmark point BP-C as we discuss below. 
Nonzero complex phases are turned on to generate a CP-asymmetry and a baryon asymmetry. 
The above study provides a concrete indication of the sizes of the
$\Chi$ decay and scattering rates and the baryon asymmetry that one can expect in our theory.
Since the baryon asymmetry from $\Chi_2$ (i.e. with $n\!=\!2$) is dominant, we restrict our analysis here to this,
and henceforth refer to it as $\Chi$, and its mass as $M$.
The mass scale at which the new physics is operative is set by $M_\chi$.
We scale all masses and rates by $M_\chi$, rendering them dimensionless, 
and work with these dimensionless quantities in our numerical analysis.
Our numerical work is in the pseudo-Dirac limit, which implies $\hat{M} \approx 1$.

\subsection{The Thermally Averaged Rates and the Hubble Expansion Rate}
\label{GmGmSigTA.SEC}

\begin{figure}
  \begin{center}
    \includegraphics[width=0.31\textwidth]{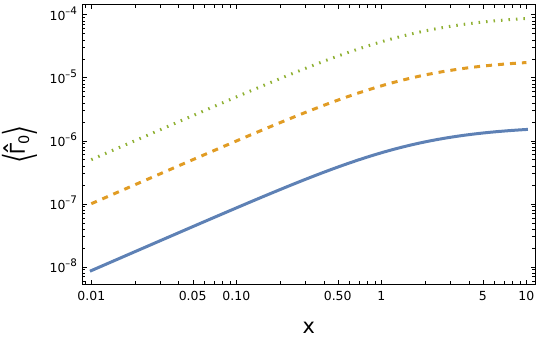} \hspace*{0.01\textwidth}
    \includegraphics[width=0.31\textwidth]{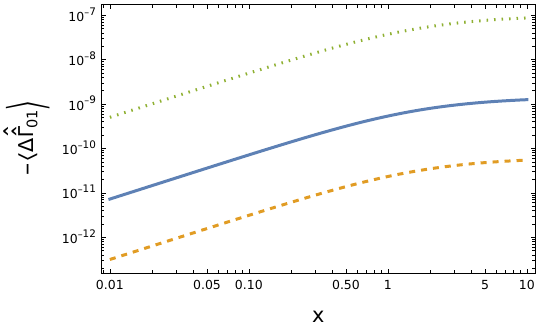} \hspace*{0.01\textwidth} \\
    \includegraphics[width=0.31\textwidth]{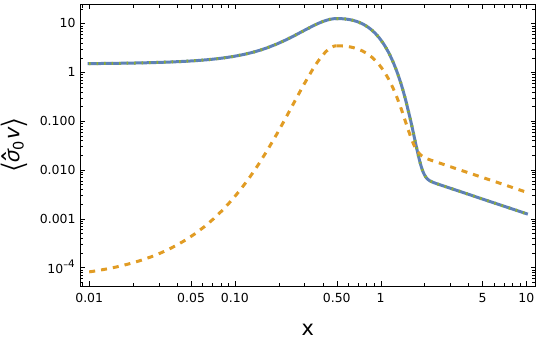} \hspace*{0.01\textwidth} 
    \includegraphics[width=0.31\textwidth]{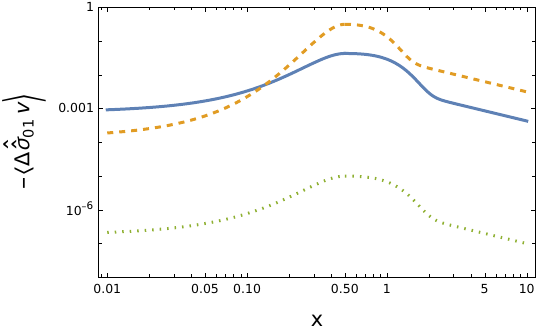} \hspace*{0.01\textwidth} 
    \includegraphics[width=0.31\textwidth]{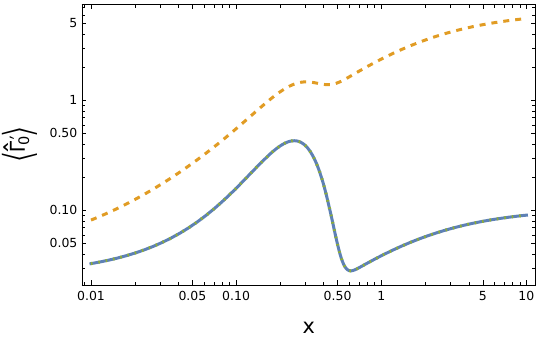}
  \end{center}    
  \caption{The dimensionless thermally averaged decay rate and difference (top row),
    and scattering cross section, difference, and $\Delta B\!=\!2$ scattering rate (bottom row),
    as a function of $x\!=\!M_\chi/T$, for benchmark scenarios BP-A (solid), BP-B (dashed), and BP-C (dotted), 
    without the $g,\rMLtxt$ scaling factors.
 \label{GmSigTAVx.FIG} }
\end{figure}

In Table~\ref{BPparams.TAB} we show a set of BPs and the choice of parameters in each BP as studied in Ref.~\cite{Gopalakrishna:2023mul}. 
We show in the first row of the Table how each dimensionless decay and scattering rate scales with $g$ and $\rMLtxt$. 
The upper table lists the BP-A, BP-B, BP-C parameters, which we collectively refer to as BP-X, with X=\{A,B,C\}. 
The BP-A and BP-B are in line with the above study,
while BP-C has been introduced in the present study to explore a different possibility of the BAU being generated in decays,
in contrast to the former two where it is mainly in scattering as we will see below.
The BP-A and BP-B values and the $x$ dependence of $\GmTA,\DGmTA,\SigvTA,\DSigvTA,\GmPTA$, are 
given in Eqs.~(\ref{GmTABPAB.EQ}), (\ref{SigTABPAB.EQ}), (\ref{GmPTABPAB.EQ}) of Appendix~\ref{GmSigvxDep.SEC}.
We set the $x$ dependence in BP-C to be as in BP-A, but with different normalizations of the decay and scattering rates as given in Table~\ref{BPparams.TAB}. 
In Fig.~\ref{GmSigTAVx.FIG}, we show the $x\!=\!M_\chi/T$ dependence of $\GmzTAh,\DGmzoTAhh,\SigvzTAh,\DSigvzoTAhh,\GmzPTAh$.
The BP-A and BP-C curves are coincident in the $\SigvzTA$ and $\GmzPTA$ plots.
The values shown in the plots are the dimensionless quantities, and are to be scaled with the $g$ and $\rMLtxt$ factors given in the first row of Table~\ref{BPparams.TAB}.
In BP-A, the extra scaling factor $(M_\chi/\Lambda)^2$ present in $\DGmzohh$, $\DSigvzoTAhh$ has not been included in the plots. 
For $x\ll 1$, the $\SigvzTAh$ is bigger in BP-A than in BP-B due to the contribution of the t-channel process
(SC-1) in the former which is not present in the latter (see Ref.~\cite{Gopalakrishna:2023mul}).  

In each of the above BP-X, we define three BPs with $M_\chi\!=\!\{10^{12},10^9,10^6\}$~GeV, 
denoting them as BP-nX for n=\{1,2,3\} respectively.
We thus have nine BPs with the mnemonic BP-nX, for n=\{1,2,3\} and X=\{A,B,C\}. 
In Table~\ref{BPparams.TAB} (lower) we show the $g$ values for each BP-nX,
and in the ``Modified parameters'' row,
we show only those parameters that do not take the standard values of the upper table, 
with the understanding that the parameters not listed in the lower table take their standard values for that BP-nX. 
\begin{table}
  \caption{The Benchmark Points (BP) parameters, and the $g,\rMLtxt$ scaling factors.
    \label{BPparams.TAB}}
  \medskip

\scriptsize  
  
\begin{centering}
\begin{tabular}{|c||c|c|c|c|c|c|}
\hline 
 & $\MQh$ & \begin{cellvarwidth}[t]
\centering
$\Gmzh$

${\footnotesize (\times|g|^{2}M_{\chi}^{4}/\Lambda^{4})}$
\end{cellvarwidth} & \begin{cellvarwidth}[t]
\centering
$\DGmzohh$

{\footnotesize$(\times{\rm Im}(g^{4})M_{\chi}^{6}/\Lambda^{6})$}
\end{cellvarwidth} & \begin{cellvarwidth}[t]
\centering
$\SigvzTAh$

{\footnotesize$(\times|g|^{2}M_{\chi}^{4}/\Lambda^{4})$}
\end{cellvarwidth} & \begin{cellvarwidth}[t]
\centering
$\DSigvzoTAhh$

{\footnotesize$(\times{\rm Im}(g^{4})M_{\chi}^{6}/\Lambda^{6})$}
\end{cellvarwidth} & \begin{cellvarwidth}[t]
\centering
$\GmzPTAh$

${\footnotesize (\times|g|^{4}\,M_{\chi}^{8}/\Lambda^{8})}$
\end{cellvarwidth}\tabularnewline
\hline 
\hline 
BP-A & $1/4$ & $1.7\times10^{-6}\,$ & $-1.4\times10^{-9}M_{\chi}^{2}/\Lambda^{2}$ & $12.7$ & $-0.04\,M_{\chi}^{2}/\Lambda^{2}$ & $0.4$\tabularnewline
\hline 
BP-B & $1/20$ & $2\times10^{-5}$ & $-6\times10^{-11}$ & $3.5$ & $-0.3$ & $6.4$\tabularnewline
\hline 
BP-C & $1/20$ & $10^{-4}$ & $-10^{-7}$ & $12.7$ & $-10^{-5}$ & $0.4$\tabularnewline
\hline 
\end{tabular}
\par\end{centering}
\medskip{}

\centering{}%
\begin{tabular}{|c||c|c|c||c|c|c||c|c|c|}
\hline 
 & BP-1A & BP-1B & BP-1C & BP-2A & BP-2B & BP-2C & BP-3A & BP-3B & BP-3C\tabularnewline
\hline 
\hline 
$M_{\chi}$ (GeV) & \multicolumn{1}{c}{} & \multicolumn{1}{c}{$10^{12}$} &  & \multicolumn{1}{c}{} & \multicolumn{1}{c}{$10^{9}$} &  & \multicolumn{1}{c}{} & \multicolumn{1}{c}{$10^{6}$} & \tabularnewline
\hline 
$M_{\chi}/\Lambda$ & $1/10$ & $1/10$ & $1/10$ & $1/10$ & $1/10$ & $1/10$ & $1/10$ & $1/50$ & $1/10$\tabularnewline
\hline 
$g$ & $0.25$ & $0.008$ & $0.048$ & $0.065$ & $0.009$ & $0.065$ & $0.024$ & $0.022$ & $0.087$\tabularnewline
\hline 
\begin{cellvarwidth}[t]
\centering
Modified

parameters
\end{cellvarwidth} &  &  &  &  &  &  & \begin{cellvarwidth}[t]
\centering
$(2/5)\,\SigvzTAh$

$(5/2)\,\DSigvzoTAhh$
\end{cellvarwidth} &  & $(1/30)\,\Gmzh$\tabularnewline
\hline 
\end{tabular}
\end{table}
These choices all give the observed BAU, i.e. $\YB(x_e)/\YB^{\rm obs}\! =\! 1$.
To get a flavor of the BAU generated in other related models that are not identical to ours but with similar ingredients,
we present our results for the peak height of the $x$ distribution (cf. Fig.~\ref{GmSigTAVx.FIG})
scaled about the values in the BP-nX, but keeping the same $x$ dependence. 

\begin{figure}
  \begin{center}
    \includegraphics[width=0.32\textwidth]{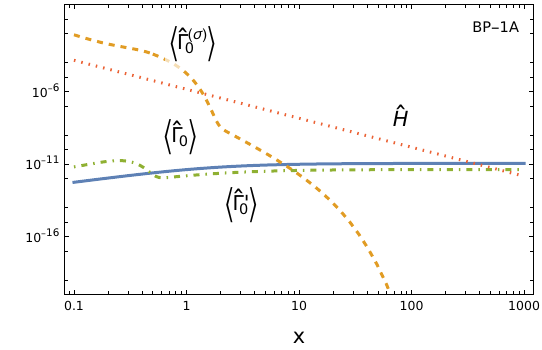}
    \includegraphics[width=0.32\textwidth]{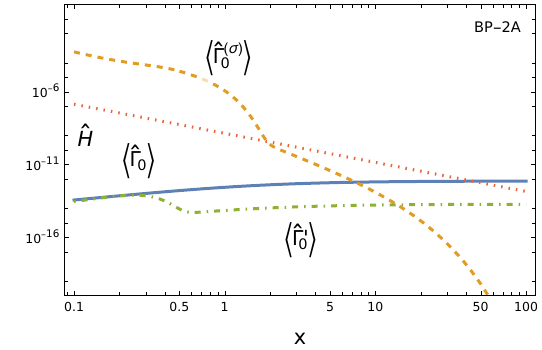}
    \includegraphics[width=0.32\textwidth]{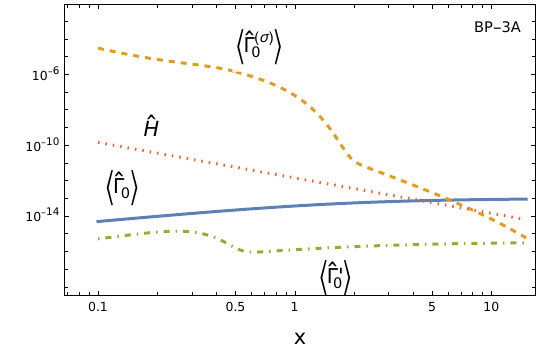}
    \\    
    \includegraphics[width=0.32\textwidth]{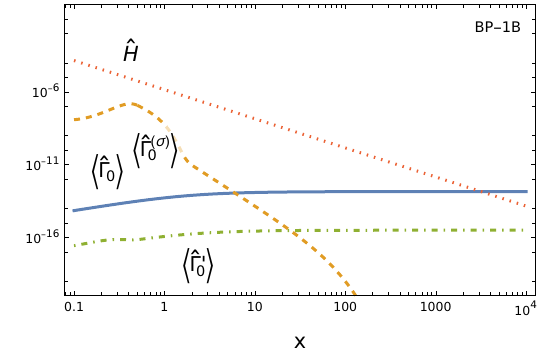}
    \includegraphics[width=0.32\textwidth]{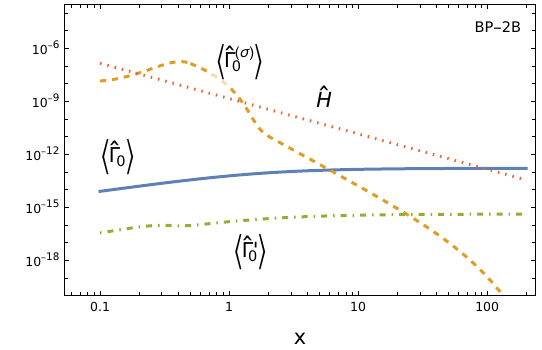}
    \includegraphics[width=0.32\textwidth]{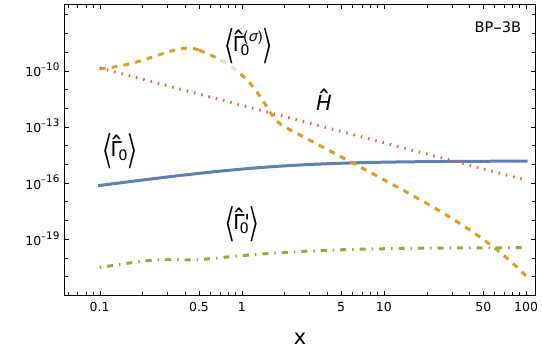}    
    \\
    \includegraphics[width=0.32\textwidth]{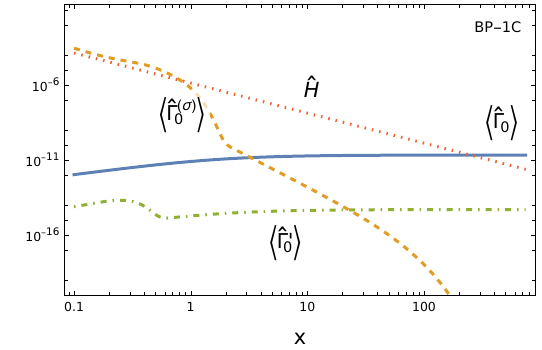}
    \includegraphics[width=0.32\textwidth]{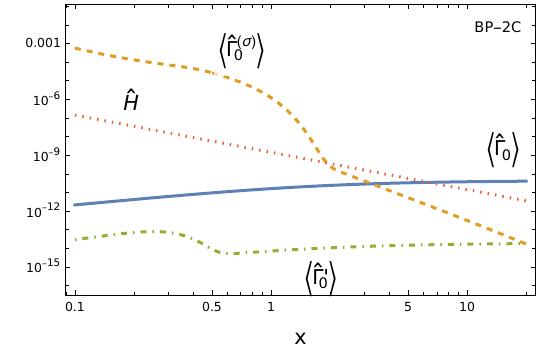}
    \includegraphics[width=0.32\textwidth]{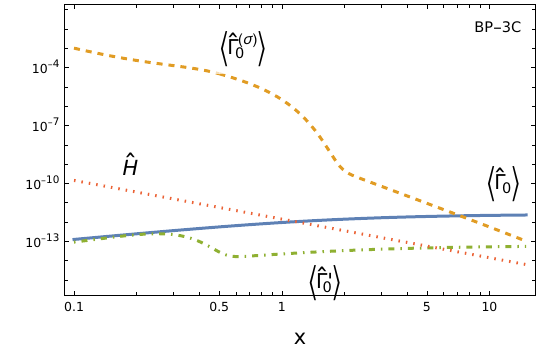}   
  \end{center}    
  \caption{The dimensionless thermally averaged decay rate $\GmzTAh$ and scattering rates $\GmzSigTAh$,~$\GmzPTAh$
    compared with the Hubble rate $\Hh$
    as a function of $x\!=\!M_\chi/T$, for benchmark points BP-1X,\,2X,\,3X (first, second, third columns), and, BP-nA,\,nB,\,nC (top, middle, bottom rows),
    where X=\{A,B,C\} and n=\{1,2,3\}, 
    with the $g,\rMLtxt$ scaling factors included. 
 \label{GmGmSigHxAB.FIG} }
\end{figure}

In Fig.~\ref{GmGmSigHxAB.FIG} we compare the dimensionless thermally averaged leading order decay and scattering rates $\GmzTA,\GmzSigTA,\GmzPTA$
(the rates on p.\,\pageref{procsDef.PG} made dimensionless by dividing by $M_\chi$)
to the dimensionless Hubble expansion rate $\Hh\!=\!H/M_\chi$, for each of the nine benchmark points BP-nX.
In all cases, we find for $x\lesssim 5$, $\GmSigTA > \GmTA,\GmPTA$,
suggesting that the scattering channels play a very important role in our framework.
For BP-nA, since $\GmSigTA\! >\! H$ for $x\!\approx\! 0.1$, the $\Chi_n$ is comfortably in TE in that epoch,
but for $x\gtrsim 1$ we have $\GmSigTA\! \lesssim\! H$ and the $\Chi_n$ falls out of TE.
For BP-nB, on the other hand, the situation is different; in BP-1B it is never in TE in the entire $x$ range of interest,
in BP-2B it starts out not being in TE, goes into, and later falls out of TE,
while in BP-3B it starts out barely being in TE, attains TE, and falls out.
In all cases (except BP-1B), comparing the BP-nX,
we see that for smaller $M_\chi$, the $\Chi_n$ is more strongly in TE at small $x$ and goes out of TE at a larger $x$. 
In BP-nA, for $x\lesssim 0.3$, $\GmPTAh$ is comparable to $\GmTAh$ but becomes somewhat smaller for larger $x$,
while in BP-nB, in the entire $x$ range $\GmPTAh$ is much smaller compared to the other rates.
$\GmPTAh$ could potentially have been very important, but given the extra $(\rMLtxt)^4$ suppression,
it ends up being either of the same order, or even suppressed, compared to $\GmTA, \GmSigTA$.

\subsection{Solution of the Boltzmann Equations and the BAU}
\label{BENumSoln.SEC}

We turn next to discussing the generation of the baryon asymmetry in the early Universe.
For each BP,
we set the thermally averaged decay and scattering rates in the BE of Eq.~(\ref{BEYchiYB.EQ})
(i.e. the collision terms in the rhs) 
to the values specified in Table~\ref{BPparams.TAB}, and the $x$ dependence as in Appendix~\ref{GmSigvxDep.SEC}, 
numerically solve the BE in {\it Mathematica~14.0},
and find for what choices of $M_\chi$, $M_\chi/\Lambda$, $g$, we obtain the observed BAU.

Our numerical solution is over $x\!=\!(x_b,x_e)$.
We take the initial conditions as $Y_\Chi(x_b) = \delta_\Chi(x_b) Y_\Chi^{(0)}(x_b)$, and $\YB(x_b) = 0$. 
If $\Chi$ is in TE at $x=x_b$, we take $\delta_\Chi(x_b) = 1$,
and if it is not, we take $\delta_\Chi(x_b)$ as we describe in Appendix~\ref{chiOutOfTEUV.SEC}
(cf. Eq.~(\ref{deltaChiPhi.EQ})).
We then numerically solve the BE, 
and from the solution, read off the final yields, $Y_\Chi(x_e)$ and $\YB(x_e)$.
We take the range to be $(x_b,x_e)\!=\!(0.1,10)$ typically, 
but if the $n_\chi$ has not depleted to a negligible value of $Y_\chi(x_e) < 10^{-4}$ and $\YB(x_e)$ has not stabilized to its asymptotic value, we increase $x_e$ to a larger value
till $\YB(x_e)$ has stabilized.
We look for regions of parameter space where $\YB(x_e)/\YBobs\! \approx \! 1$.

\begin{figure}
  \begin{center}
    \includegraphics[width=0.3\textwidth]{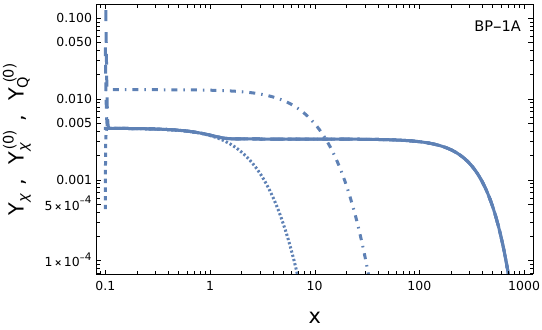}\hspace*{0.03\textwidth}
    \includegraphics[width=0.3\textwidth]{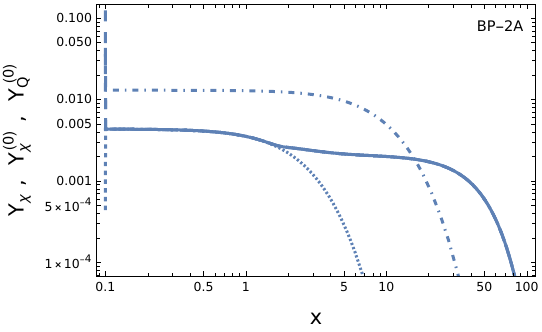}\hspace*{0.03\textwidth}
    \includegraphics[width=0.3\textwidth]{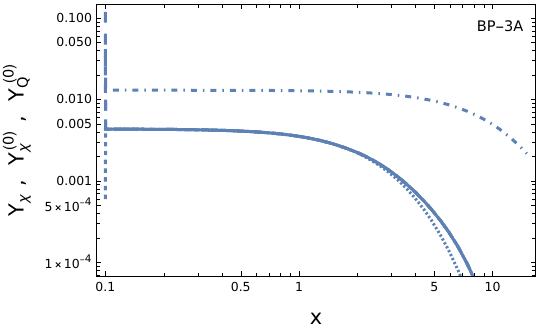}\hspace*{0.03\textwidth}    
    \\   
    \includegraphics[width=0.3\textwidth]{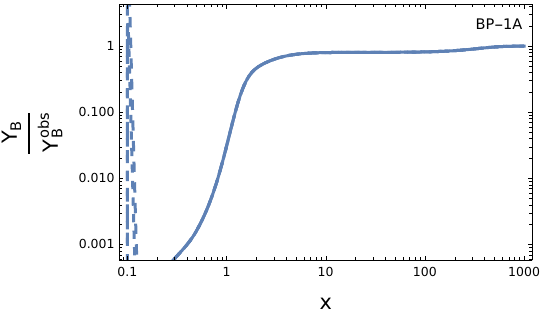}\hspace*{0.03\textwidth}
    \includegraphics[width=0.3\textwidth]{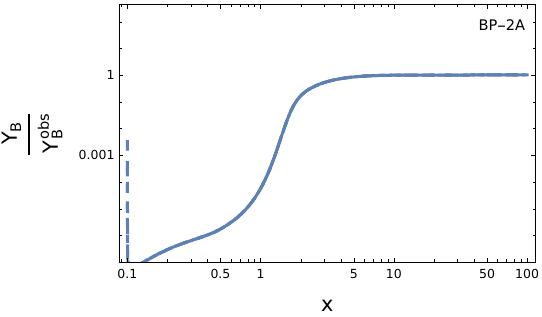}\hspace*{0.03\textwidth}
    \includegraphics[width=0.3\textwidth]{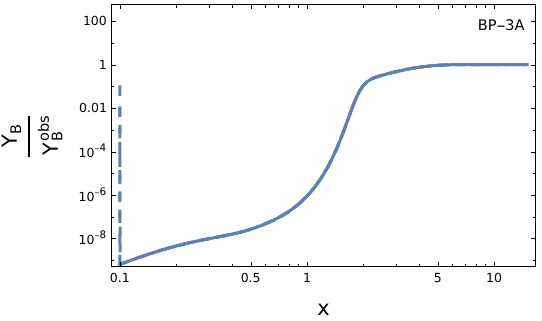}\hspace*{0.03\textwidth}
    \\
\vspace*{0.5cm}
    \includegraphics[width=0.3\textwidth]{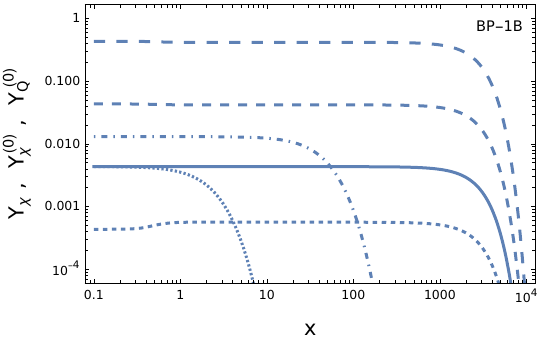}\hspace*{0.03\textwidth}
    \includegraphics[width=0.3\textwidth]{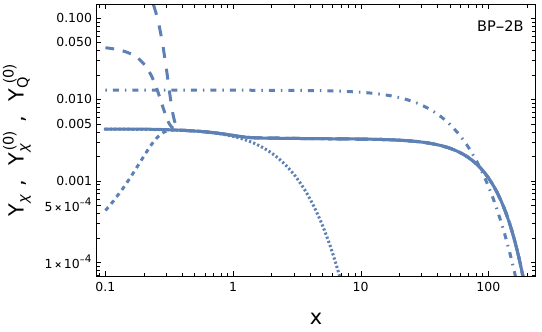}\hspace*{0.03\textwidth}
    \includegraphics[width=0.3\textwidth]{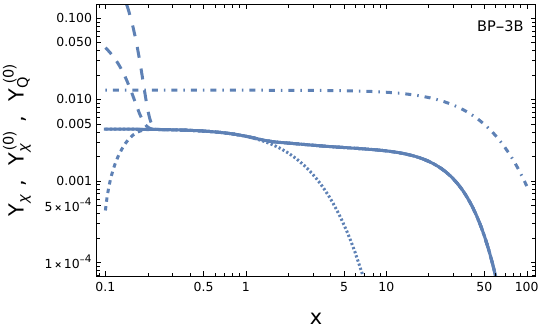}\hspace*{0.03\textwidth}
    \\
    \includegraphics[width=0.3\textwidth]{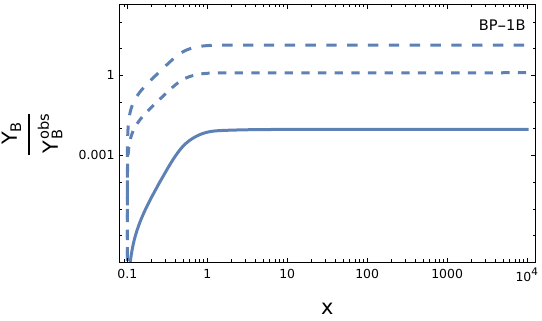}\hspace*{0.03\textwidth}
    \includegraphics[width=0.3\textwidth]{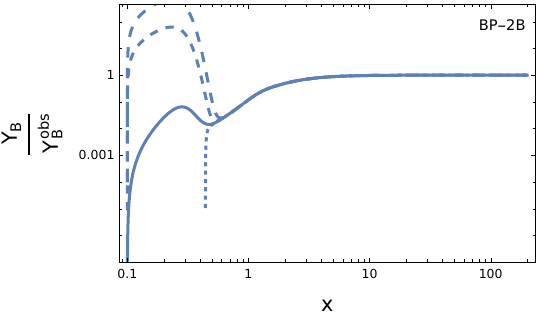}\hspace*{0.03\textwidth}
    \includegraphics[width=0.3\textwidth]{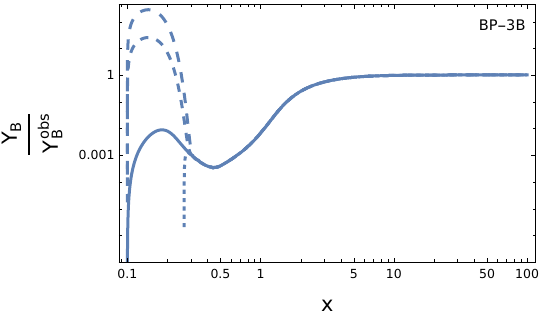}\hspace*{0.03\textwidth}    
    \\
\vspace*{0.5cm}
    \includegraphics[width=0.3\textwidth]{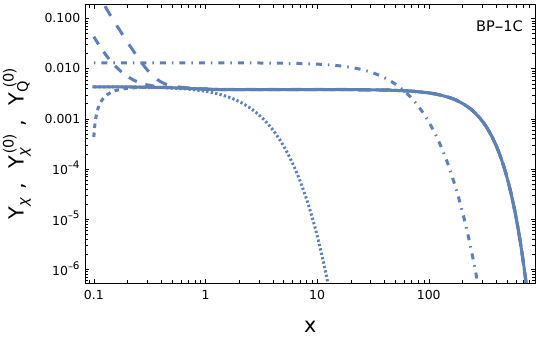}\hspace*{0.03\textwidth}
    \includegraphics[width=0.3\textwidth]{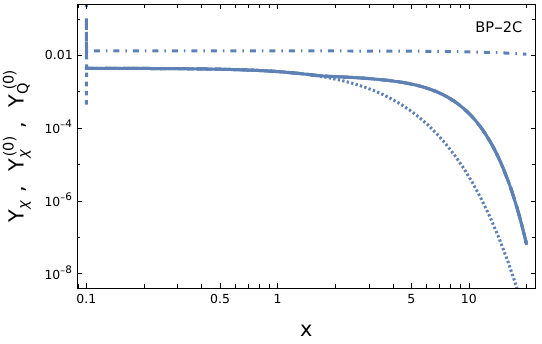}\hspace*{0.03\textwidth}
    \includegraphics[width=0.3\textwidth]{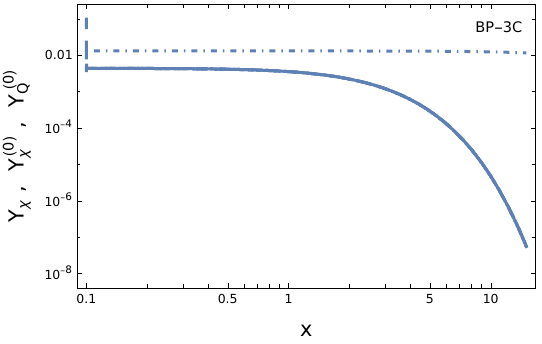}\hspace*{0.03\textwidth}
    \\
    \includegraphics[width=0.3\textwidth]{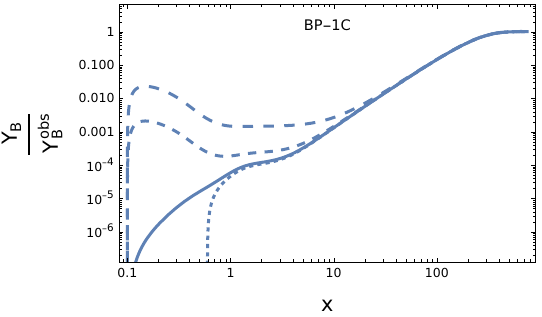}\hspace*{0.03\textwidth}
    \includegraphics[width=0.3\textwidth]{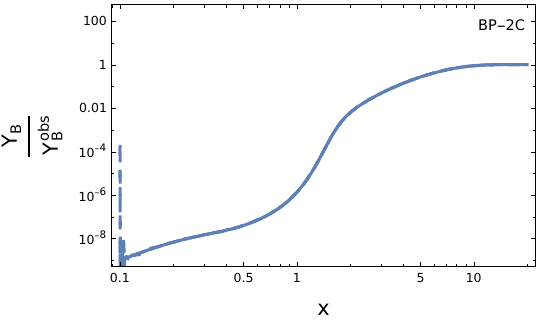}\hspace*{0.03\textwidth}
    \includegraphics[width=0.3\textwidth]{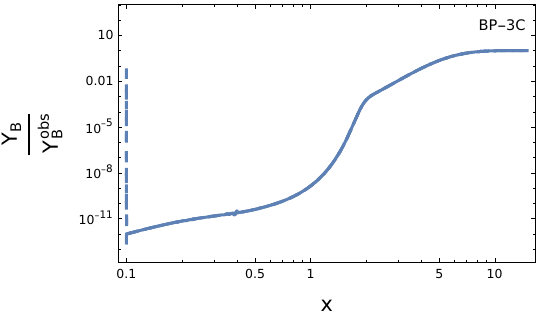}\hspace*{0.03\textwidth}        
  \end{center}    
  \caption{
    For the benchmark points BP-1X,\,2X,\,3X (left, middle, right columns),
    BP-nA,\,nB,\,nC (top-two, middle-two, last-two rows),  
    the $Y_\Chi(x)$ and $\YB(x)/\YBobs$, for $\delta(x_b) = 1$ (solid curve) and $\delta(x_b) = \{0.1,10,100\}$ (dashed curves with increasing dash-lengths),
    the dotted curve showing $\Ychiz(x)$, and, the dot-dashed curve showing $\YQz(x)$.
 \label{YGmHdelSmp123AB.FIG} } 
\end{figure}

We discuss next a few salient features of the solutions we obtain upon numerical integration of the BE. 
In Fig.~\ref{YGmHdelSmp123AB.FIG}
we show the numerical solutions of the BE for $\Ychi(x)$ along with the equilibrium $\Ychiz(x)$ and $\YQz(x)$, 
and $\YB(x)/\YBobs$,
for BP-nA,\,nB,\,nC (top-two, middle-two, and bottom-two rows).    
We see from the $\YB(x)/\YBobs$ plots how the baryon number yield builds up as the Universe cools,
i.e. as $x\!=\!M_\chi/T$ increases.
We see that our choice of the parameters in each BP indeed results in $\YB(x)\!=\!\YBobs$.
Most of the $\YB$ builds up in the range $x\in (0.1,10)$, except in BP-1C when it is over $(100,1000)$. 
We find $\delta_\Chi(x_b)\! =\! 1$ for BP-nA,\,nC since the $\Chi_n$ is in TE at $x_b$. 
In BP-nB, the $\Chi$ is not in TE at $x_b$, and we fix the initial overdensity of $\Chi$
following Appendix~\ref{chiOutOfTEUV.SEC},  
which gives $\delta_\Chi(x_b)\! =\! (8, 8, 3.3)$ for $n=1,2,3$. 
To cover other UV completion possibilities fixing the $\delta_\Chi(x_b)$ differently, we show other choices also,
namely $\delta_\Chi(x_b)\!=\! (0.1,1,10,100)$, for all BP-nX.
In all cases (except BP-1B), even if the $\Chi$ starts out not being in TE at $x_b$, it quickly thermalizes; only in BP-1B it never does.
We see that in all cases except BP-1B, the final $\YB(x_e)$ is independent of the choice of $\delta_\Chi(x_b)$ since $\Ychi$ is driven toward TE
as evidenced by the dashed lines merging with the solid curve; 
in BP-1B, since the $\Ychi$ remains out of TE, the final $\YB(x_e)$ depends on the initial $\delta_\Chi(x_b)$
as we would expect; for our choice of parameters the observed $\YB$ is obtained for $\delta_\Chi(x_b) = 8$. 
Thus, the final $\YB$ is independent of the UV completion details in all cases, except in BP-1B.  
Comparing the $\Ychi(x)$ with the $\YSMz\! \approx\! 0.23$ (cf. Appendix~\ref{ThDUnivRev.SEC}), we find that
in some cases, although most of the $\YB$ is generated for $x < 10$ (except in BP-1C),
a substantial $\Chi$ fraction may remain for $x>10$, and $\Ychi(x_e)/\YSMz\!$ falls to $< 10^{-4}$ only for much larger $x$. 
In BP-nA,\,nB, in which the baryon asymmetry is generated in scattering,
we see that the $\YB$ builds up when $\GmSigTA \lesssim H$ (cf. Fig.~\ref{GmGmSigHxAB.FIG}), 
while in BP-nC since it is generated in decay, it builds up when $\GmTA \lesssim H$,
and is particularly delayed in BP-1C to $x\gtrsim 100$. 
In BP-2B,\,3B, we see that as $\GmSigTA$ rises above $H$ for $x\gtrsim 0.1$ the $\YB(x)$ that has accumulated by then is washed out substantially,
and when $\GmSigTA$ falls below $H$, the $\YB(x)$ builds up again and reaches its final value.

\begin{figure}
  \begin{center}
    \includegraphics[height=0.125\textheight]{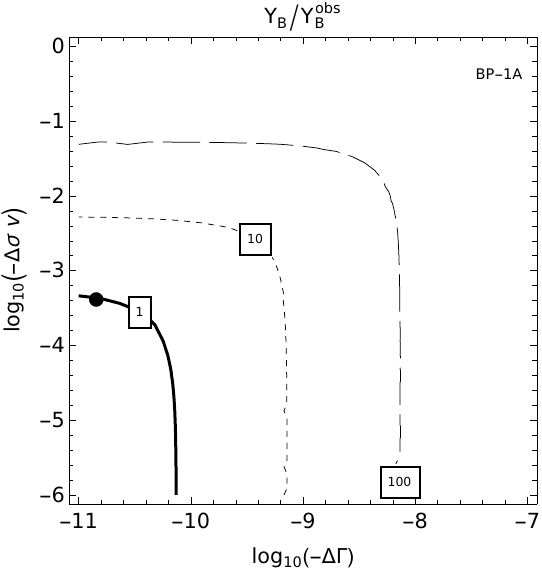}\hspace*{0.03\textwidth}
    \includegraphics[height=0.125\textheight]{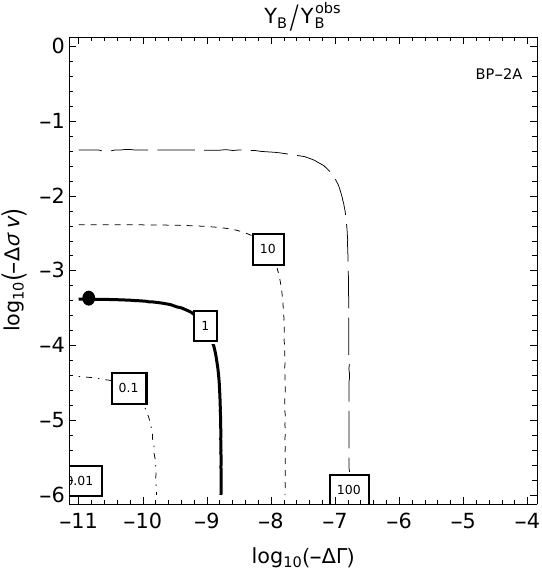}\hspace*{0.03\textwidth}
    \includegraphics[height=0.125\textheight]{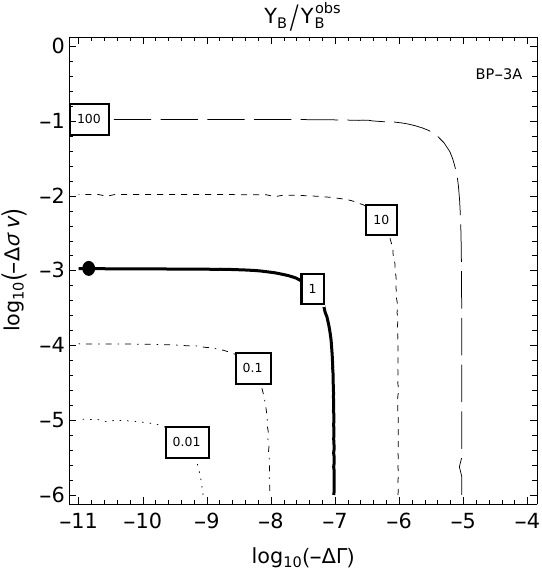}\hspace*{0.03\textwidth}    
    \\
    \includegraphics[height=0.125\textheight]{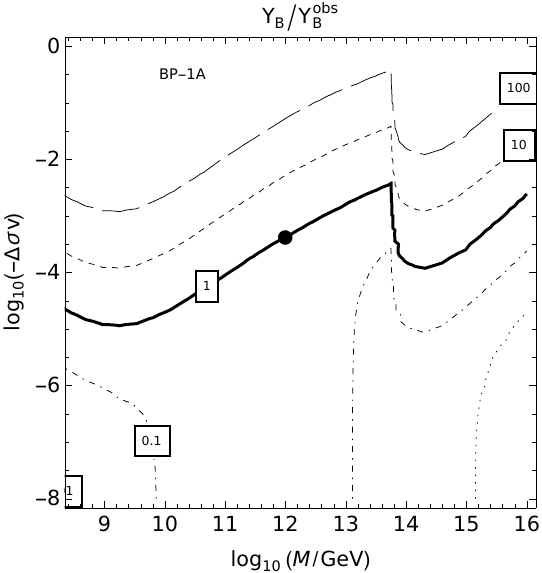}\hspace*{0.03\textwidth}
    \includegraphics[height=0.125\textheight]{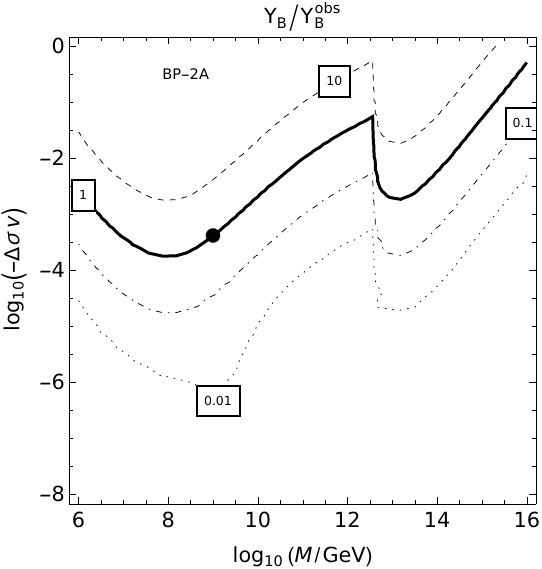}\hspace*{0.03\textwidth}
    \includegraphics[height=0.125\textheight]{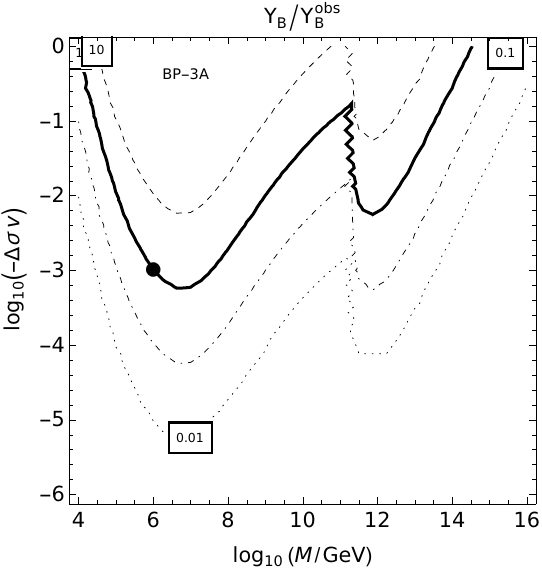}\hspace*{0.03\textwidth}
    \\
    \vspace*{1cm}
    \includegraphics[height=0.125\textheight]{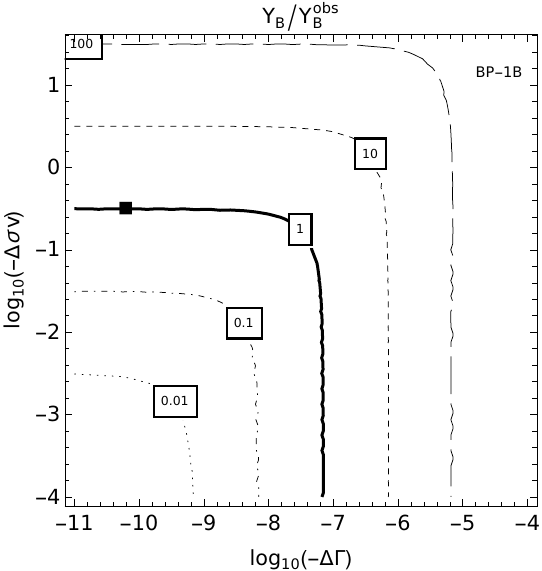}\hspace*{0.03\textwidth}
    \includegraphics[height=0.125\textheight]{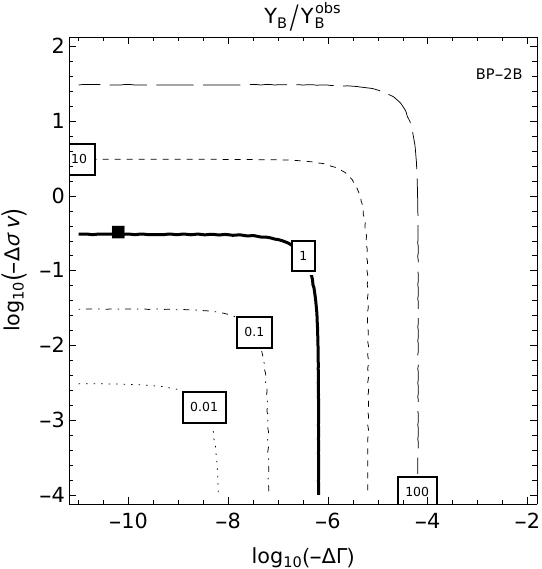}\hspace*{0.03\textwidth}
    \includegraphics[height=0.125\textheight]{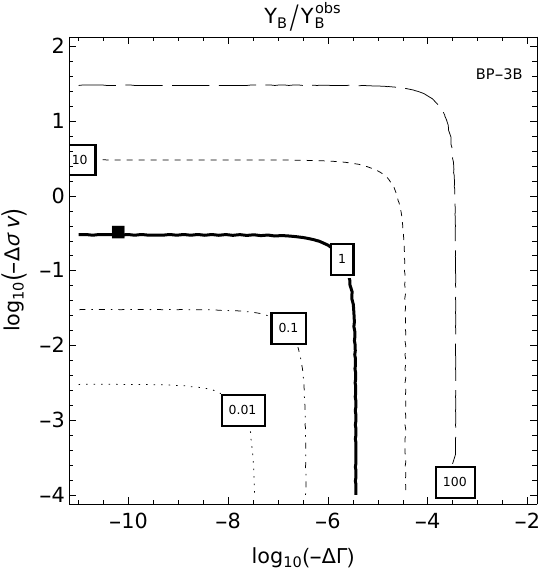}\hspace*{0.03\textwidth}
    \\
    \includegraphics[height=0.125\textheight]{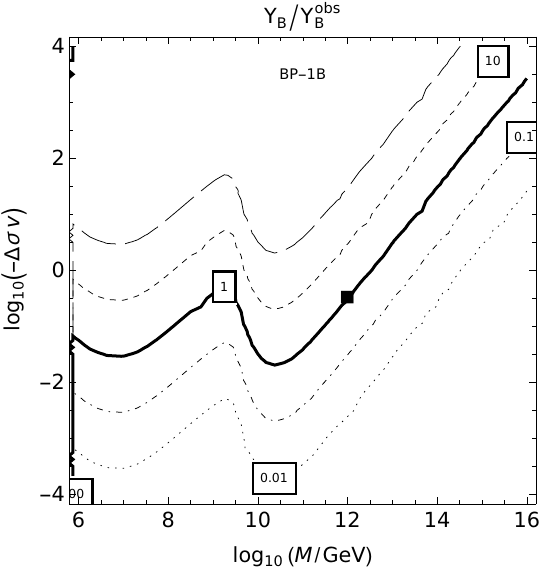}\hspace*{0.03\textwidth}
    \includegraphics[height=0.125\textheight]{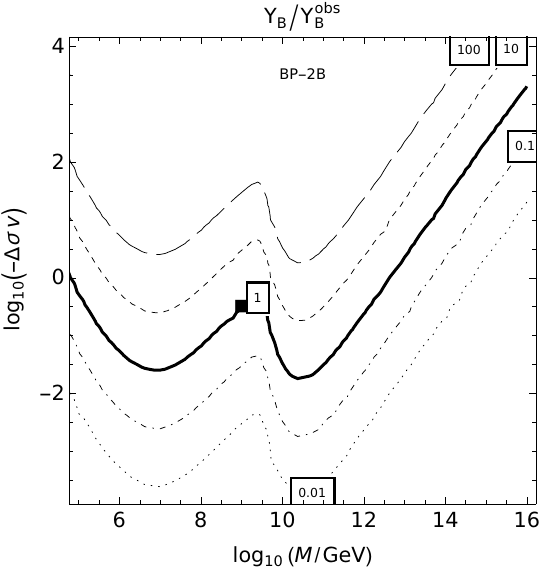}\hspace*{0.03\textwidth}
    \includegraphics[height=0.125\textheight]{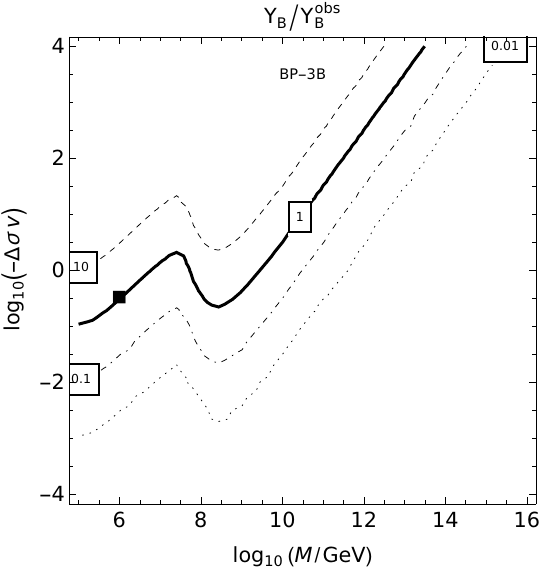}\hspace*{0.03\textwidth}
    \\
    \vspace*{1cm}
    \includegraphics[height=0.125\textheight]{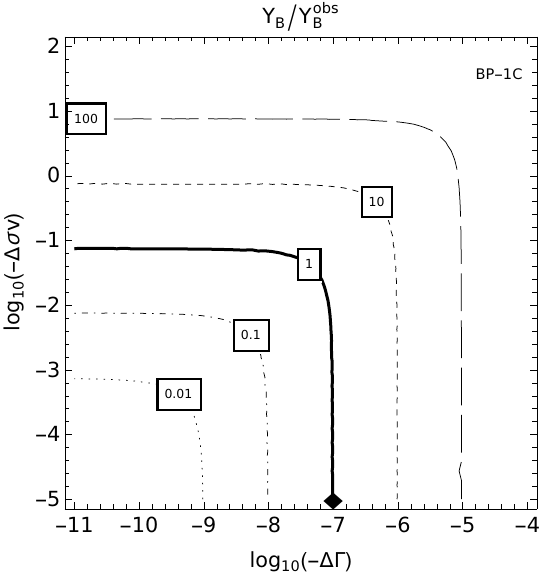}\hspace*{0.03\textwidth}
    \includegraphics[height=0.125\textheight]{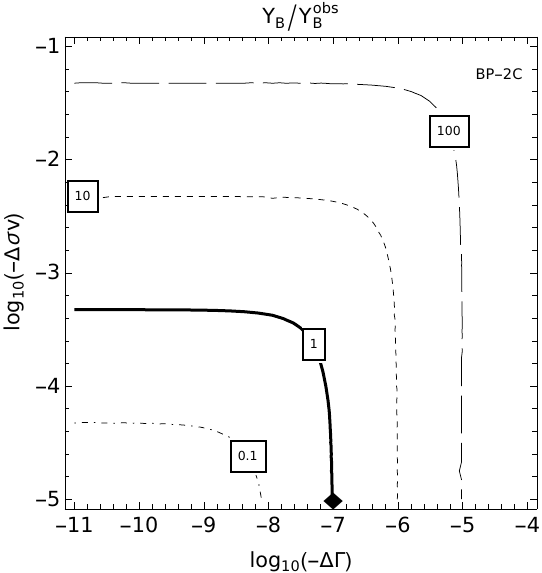}\hspace*{0.03\textwidth}
    \includegraphics[height=0.125\textheight]{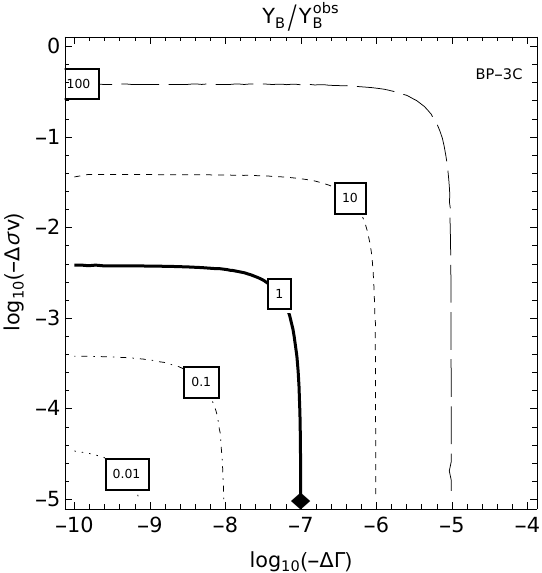}\hspace*{0.03\textwidth}
    \\
    \includegraphics[height=0.125\textheight]{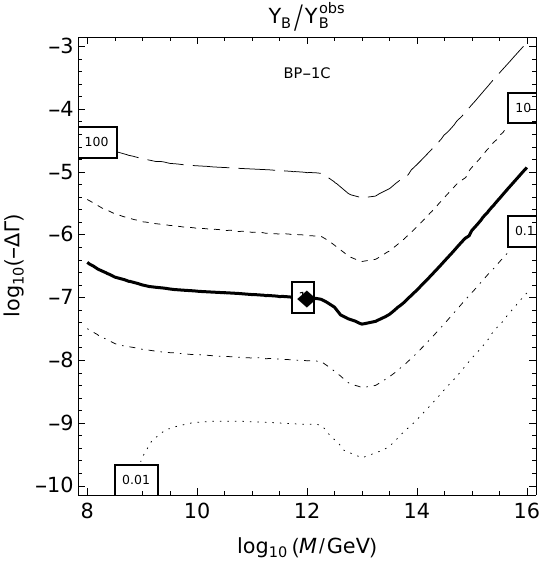}\hspace*{0.03\textwidth}
    \includegraphics[height=0.125\textheight]{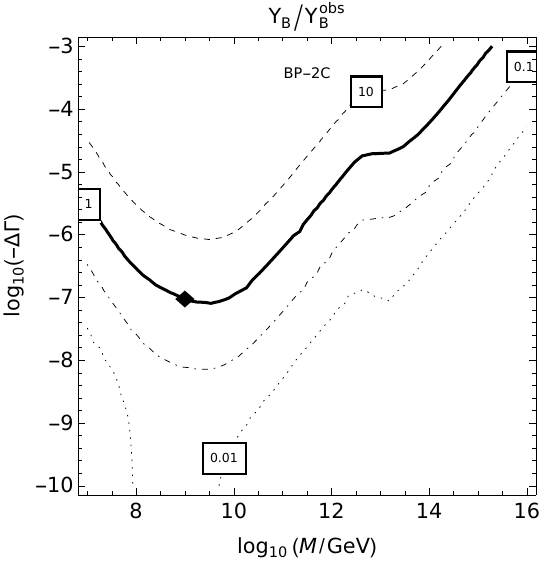}\hspace*{0.03\textwidth}
    \includegraphics[height=0.125\textheight]{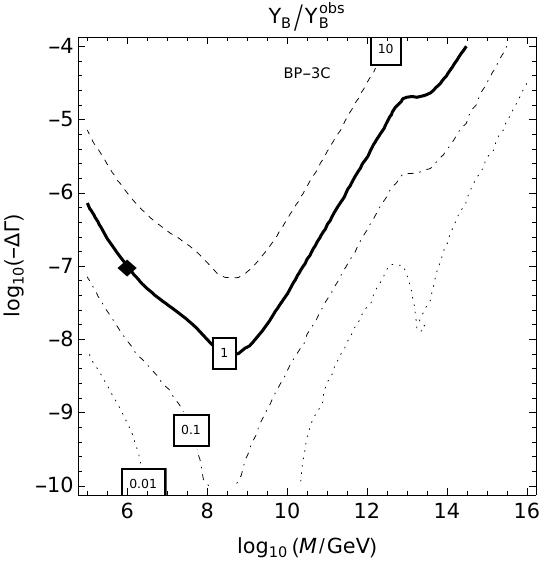}\hspace*{0.03\textwidth}    
  \end{center}    
  \caption{Contours of the final $\YB/\YB^{\rm (obs)}$ for variations about the benchmark points
    BP-1X,\,2X,\,3X (left, middle, right columns),
    and BP-nA,\,nB,\,nC (top-two, middle-two, bottom-two rows),
    with only the parameters along the axes varied.
    The benchmark point values are shown as the solid dots (for BP-nA), squares (for BP-nB), and diamonds (for BP-nC). 
 \label{YBvGmSigSmp123AB.FIG}}
\end{figure}

\begin{figure}
  \begin{center}
    \includegraphics[width=0.3\textwidth]{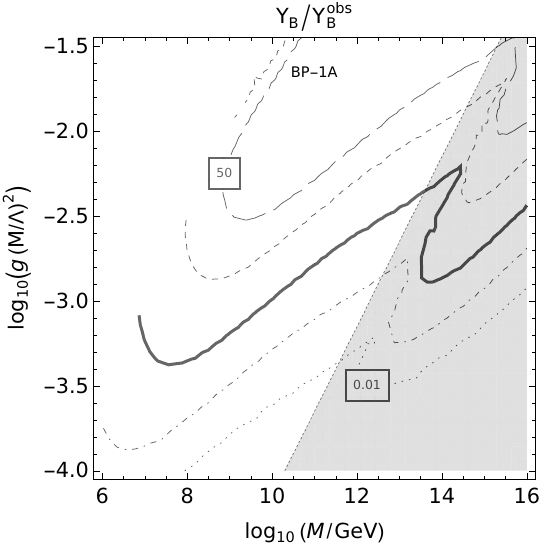}\hspace*{0.03\textwidth}    
    \includegraphics[width=0.3\textwidth]{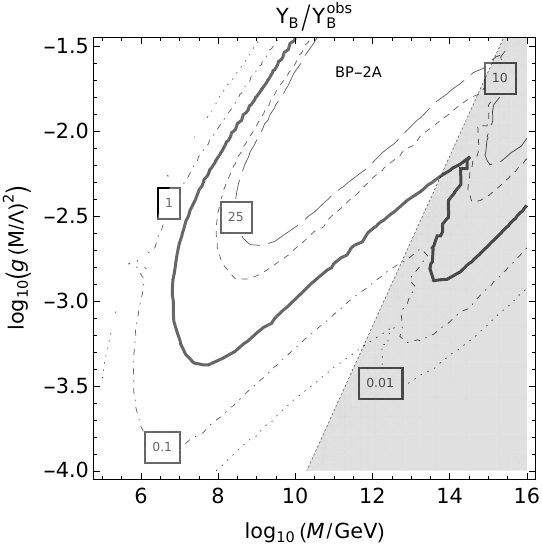}\hspace*{0.03\textwidth}
    \includegraphics[width=0.3\textwidth]{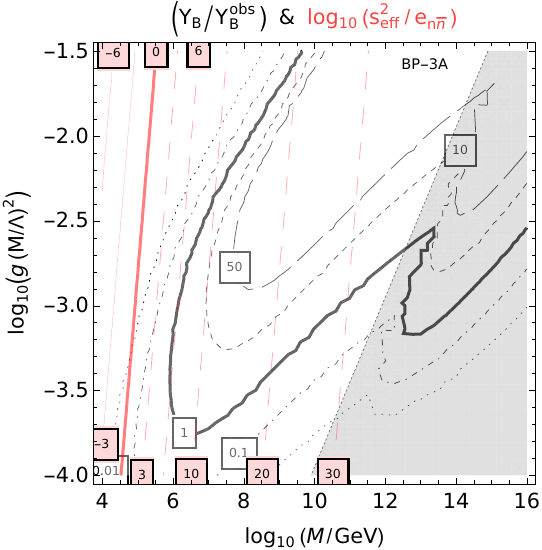}\hspace*{0.03\textwidth}
    \\
    \vspace*{0.5cm}
     \includegraphics[width=0.3\textwidth]{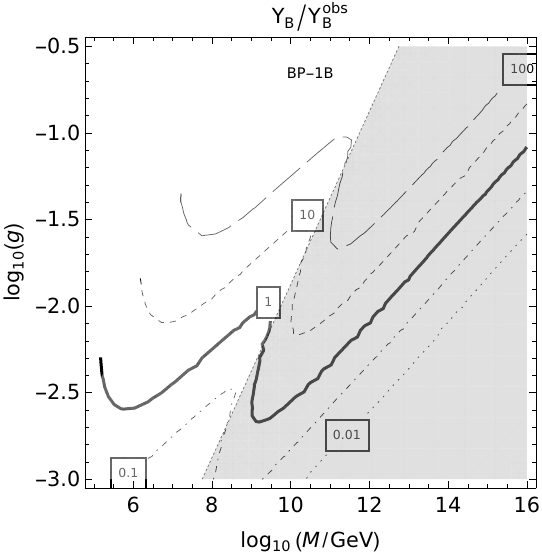}\hspace*{0.03\textwidth}
    \includegraphics[width=0.3\textwidth]{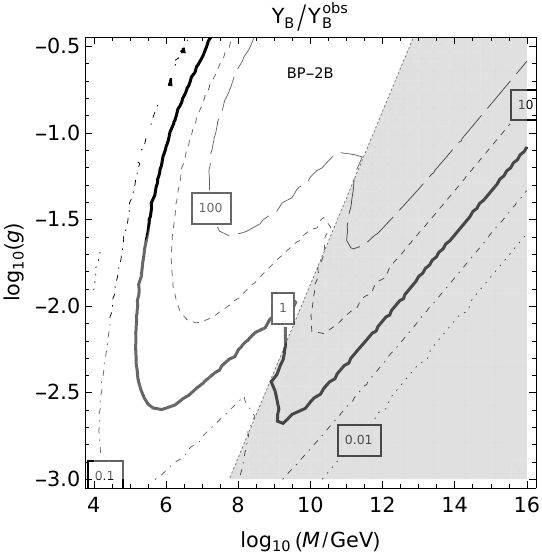}\hspace*{0.03\textwidth}
    \includegraphics[width=0.3\textwidth]{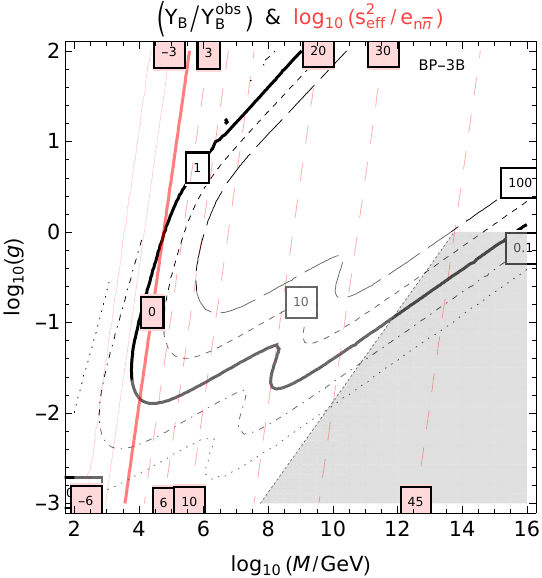}\hspace*{0.03\textwidth}
    \\
    \vspace*{0.5cm}
    \includegraphics[width=0.3\textwidth]{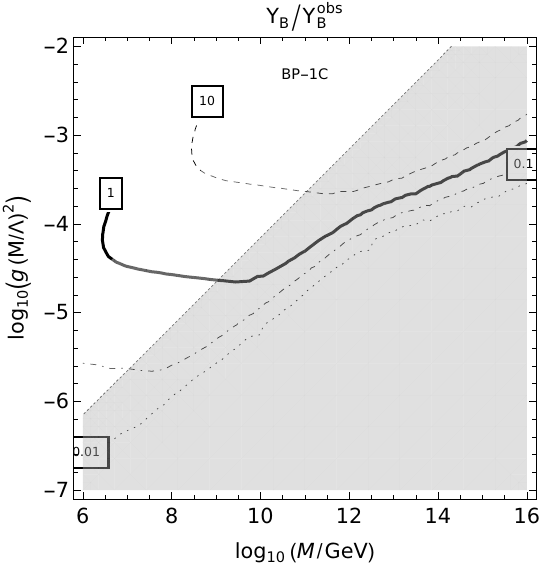}\hspace*{0.03\textwidth}    
    \includegraphics[width=0.3\textwidth]{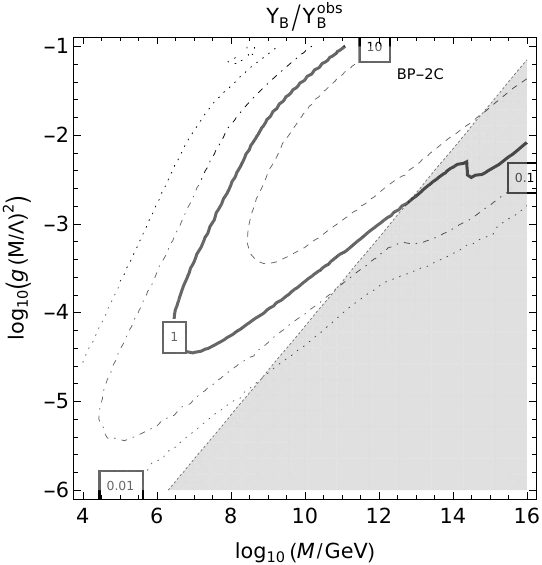}\hspace*{0.03\textwidth}
    \includegraphics[width=0.3\textwidth]{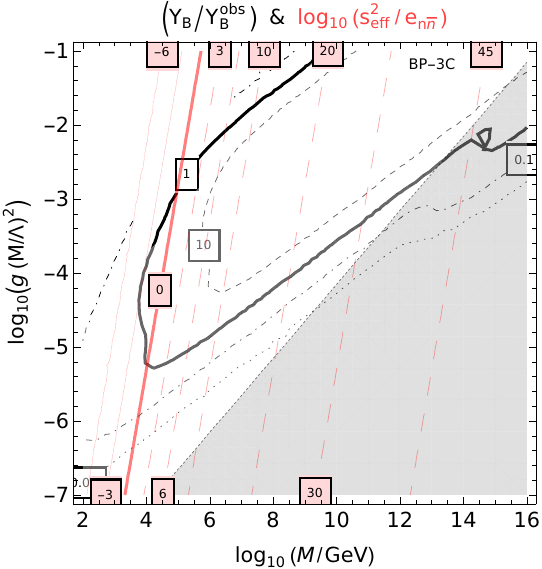}\hspace*{0.03\textwidth}      
  \end{center}    
  \caption{Contours of the final $\YB/\YB^{\rm (obs)}$ (black lines) for the benchmark points
    BP-1X,\,2X,\,3X (left, middle, right columns),
    and BP-nA,\,nB,\,nC (top, middle, bottom rows),
    with only the parameters along the axes varied.
    The light shaded region is where the $\Chi$ is decoupled at $x=1$.
    Contours of $\kappa_\nnBmm\!=\!\log_{10}{(\seff^2/e_\nnBmm)}$ (red slanted lines) show the \nnbar\ oscillation reach of current and future experiments, 
    with the region to the left of the thick red line already being nontrivially constrained by the present experimental bound. 
 \label{YBVgMrMLSmp123AB.FIG} }
\end{figure}

In Fig.~\ref{YBvGmSigSmp123AB.FIG}
we show for BP-nA, BP-nB, BP-nC the contours of the final baryon number yield
(i.e. $\YB(x_e)/\YB^{\rm (obs)}$)
varying two of the $\DGmzoTAhh,\DSigvzoTAhh,M_\chi$ 
parameters at a time about the BP values shown as the solid dots. 
The values shown along the axes correspond to rescaled peak values (including the $(M_\chi/\Lambda)^2$ factor for BP-nA)
of Eqs.~(\ref{GmTABPAB.EQ}),~(\ref{SigTABPAB.EQ}),~(\ref{GmPTABPAB.EQ}), with the $x$ functional form in those kept intact.
The parameters not varied are all fixed at the values for the corresponding BP. 
It is clear from the $\DGm-\DSigv$ plots that in BP-nA and BP-nB for the parameters in Table~\ref{BPparams.TAB} marked by the solid dots,
the baryon asymmetry is being generated
in $\Chi_n$ scattering processes with very little from $\Chi_n$ decay,
while in BP-nC it is the opposite. 
In Fig.~\ref{YBVgMrMLSmp123AB.FIG}
we show
contours of $\YB(x_e)/\YB^{\rm (obs)}$ varying $M_\chi$, $g$, $M_\chi/\Lambda$, two at a time. 
We show as the light shaded region where the $\Chi$ is decoupled, i.e. $\GmSigTA \leq H$, at $x=1$.
In the figure, we also overlay the neutron-antineutron oscillation rate contours as red slanted lines,
which we explain in Sec.~\ref{nnbarOsc.SEC}. 

\begin{figure}
  \begin{center}
    \includegraphics[width=0.3\textwidth]{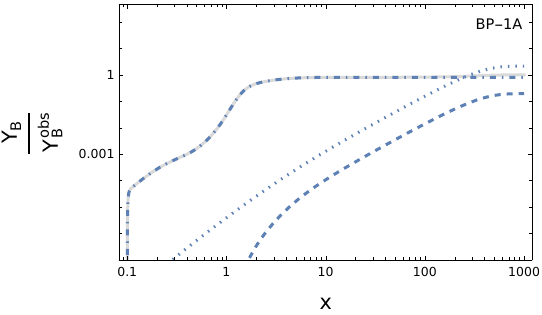}\hspace*{0.03\textwidth}
    \includegraphics[width=0.3\textwidth]{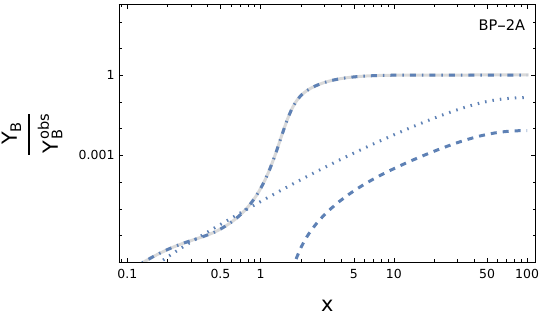}\hspace*{0.03\textwidth}
    \includegraphics[width=0.3\textwidth]{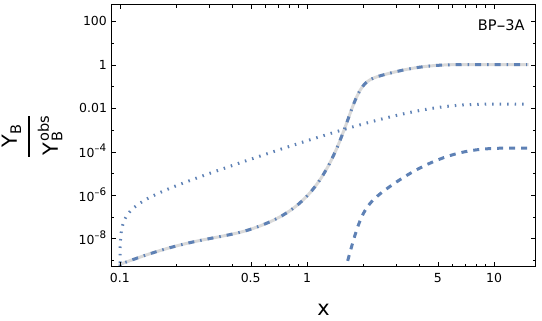}\hspace*{0.03\textwidth}
    \\
    \includegraphics[width=0.3\textwidth]{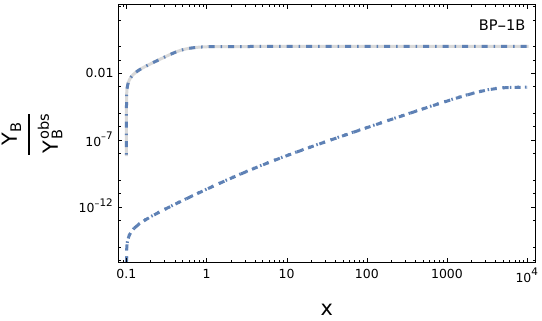}\hspace*{0.03\textwidth}
    \includegraphics[width=0.3\textwidth]{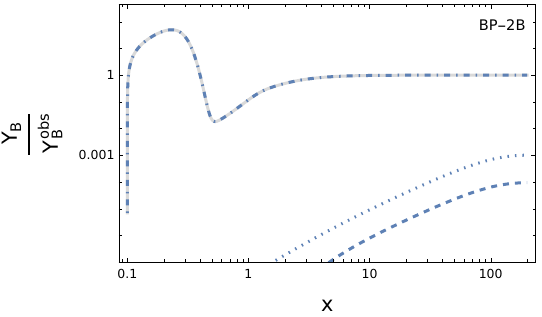}\hspace*{0.03\textwidth}
    \includegraphics[width=0.3\textwidth]{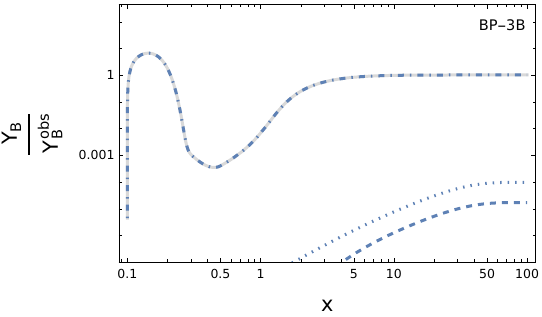}\hspace*{0.03\textwidth}
    \\
    \includegraphics[width=0.3\textwidth]{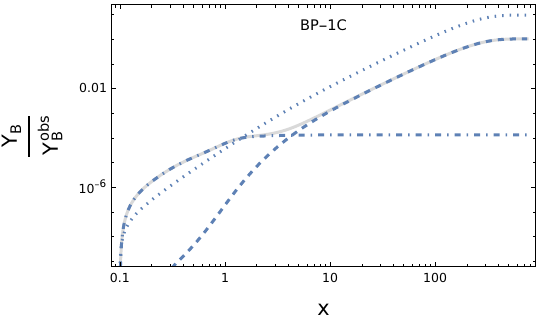}\hspace*{0.03\textwidth}
    \includegraphics[width=0.3\textwidth]{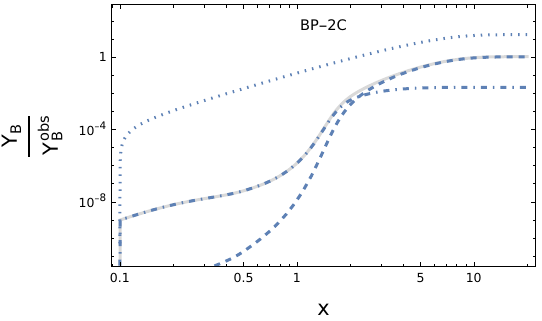}\hspace*{0.03\textwidth}
    \includegraphics[width=0.3\textwidth]{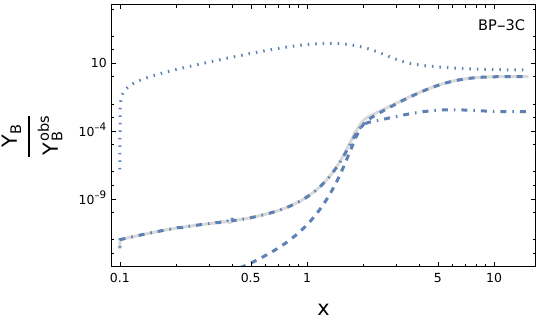}\hspace*{0.03\textwidth}
  \end{center}    
  \caption{The final $\YB(x_e)/\YB^{\rm (obs)}$ for the benchmark points
    BP-1X,\,2X,\,3X (left, middle, right columns),
    for BP-nA,\,nB,\,nC (top, middle, bottom rows),
    with all contributions included (light-gray), 
    from decay asymmetry only i.e. $\DSigv=0$ (dashed), scattering asymmetry only i.e. $\DGm=0$ (dot-dashed),
    and, only decay and no scattering i.e. $\Sigvz=0$, $\DSigv=0$ (dotted). 
 \label{YBovYBobsDS.FIG} }
\end{figure}

In Fig.~\ref{YBovYBobsDS.FIG} we show a breakup of the baryon asymmetry generated in scattering and decay processes.
As already mentioned, we see that the baryon asymmetry is generated in scattering processes in BP-nA and BP-nB,
since the dash-dot curve almost coincides with the solid curve.
The dashed curve shows that the decay process generates a very small portion of the baryon asymmetry in BP-nA and BP-nB,
while in BP-nC it is a large portion.
The fact that the dashed curve is significantly below the dotted curve (except in BP-1B) tells us that if scattering is present without any scattering asymmetry,
it washes out quite a large fraction of the baryon asymmetry generated in decay.
In BP-nC, it is the decay process that plays the dominant role, with scattering being sub-dominant. 

Our main inference from these results is that
the observed baryon asymmetry is obtained for masses in a very wide range $M_\chi\!\sim\! (10^6,10^{16})$~GeV in BP-nA,
and $M_\chi\!\sim\! (10^4,10^{16})$~GeV in BP-nB and BP-nC.
The lower mass scales are very interesting from the point of view of being able to be probed in upcoming experiments, 
such as neutron-antineutron oscillation, which we discuss next. 

\subsection{Terrestrial Probes: Neutron-Antineutron (\nnbar) Oscillation}
\label{nnbarOsc.SEC}

Having identified the mass scale $M_\chi$ and the corresponding $g$, $\rMLtxt$ values for which 
the observed BAU is obtained in our theory, 
we contemplate here terrestrial experiments that could help us test this new physics.

Proton decay is forbidden in our theory at the perturbative level because Lorentz invariance requires at least one final state fermion,
either a lepton or antilepton as they are the only SM fermions lighter than the proton, 
and this is not allowed in our theory since $L$ is a good symmetry. 

The $\Chi$ induces $\Delta B\! =\! 2$ processes,
for instance, the neutron-antineutron (\nnbar) oscillation;
see Refs.~\cite{Kuzmin:1970nx,Glashow:1979nm,Mohapatra:1980qe}, Refs.~\cite{Mohapatra:2009wp,Phillips:2014fgb} for reviews,
and Refs.~\cite{Babu:2006xc,Babu:2013yca,Grojean:2018fus,Fridell:2021gag} for some recent studies. 
The \nnbar\ oscillation is being searched for in experiments but no positive signal has been detected to date. 
Currently, among the strongest bounds on the oscillation time is 
from the Super-Kamiokande~(SK) experiment, which is $\hat\tau_{n\bar n}\geq 4.7\times 10^8$~s~\cite{Super-Kamiokande:2020bov} at the 90\% confidence limit,
which implies $({\Delta \hat{m}}_{n\bar{n}})_{\rm SK} = 1/\hat\tau_{n\bar n} \leq 10^{-34}$~GeV.
Ref.~\cite{FileviezPerez:2022ypk} summarizes the theoretical landscape and the prospects in future experiments; 
for instance, the HIBEAM/NNBAR experiment at the European Spallation Source (ESS)
anticipates a three orders of magnitude improvement~\cite{Addazi:2020nlz} in the sensitivity.

In our theory, starting from Eq.~(\ref{LChiSMudd.EQ}) and integrating out the $\Chi$ we obtain 
an effective operator~\cite{Gopalakrishna:2022hwk}
$$\QVV \equiv \epsilon^{abc}\epsilon^{a'b'c'} (\widebar{u^c_{c'}} \gamma^\nu\gamma^\mu P_{\sssty L,R}\, u_c)\, (\widebar{d^c_{b}} \gamma_\mu d_a)\, (\widebar{d^c_{b'}} \gamma_\nu d_{a'}) \ ,$$
where the subscript on the fields and on the $\epsilon$ tensor denote color indices,
the superscript $c$ on the fields denotes charge-conjugation, 
and we call attention to the vector nature of this operator.
Clearly, this operator induces neutron-antineutron oscillations. 
The \nnbar\ oscillation rate in our theory is discussed in detail in Ref.~\cite{Gopalakrishna:2022hwk} and can be written as  
\beq
{\Delta m}_{n\bar{n}} = \frac{g^2 \seff^2}{\Lambda^4 M_\chi} \matel{\QVV}{\bar{n}}{n} \ ,
\label{DmNNB.EQ} 
\eeq
where $\seff$ is as discussed in Sec.~\ref{BGChiTh.SEC}.  
 
The matrix elements $\matel{Q_i}{\bar{n}}{n} \equiv \LamQCD^6 \langle \hat{Q}_i \rangle$ are most reliably computed using lattice gauge theory (LGT) techniques. 
Since, to the best of our knowledge, there is no LGT computation available for vector operators as appropriate for us,
we Fierz rearrange the $\QVV$ operator as explained in Ref.~\cite{Gopalakrishna:2022hwk} to obtain related scalar operators (plus other vector and tensor operators),
whose matrix elements have been computed in LGT, for example, in Ref.~\cite{Rinaldi:2018osy}.
In the basis of scalar operators there, the $Q_3,Q_5,Q_7$ are relevant to the Fierz transformed $\QVV$ operator, with $Q_3$ numerically the largest,
being $\langle \hat{Q}_3 \rangle = -13$ in our notation, for $\LamQCD = 0.18$~GeV. 
Till such time that LGT results for the vector operator matrix elements become available,  
we content ourselves to presenting our results by estimating the oscillation rate
taking this scalar operator matrix element obtained in the Fierz rearrangement. 

We wish to derive next the \nnbar\ oscillation bound on our theory from SK,
and also infer the future prospects in upcoming experiments.
For this, we define the improvement factor $e_\nnBmm$ of an experiment's sensitivity relative to SK, 
and write its sensitivity as ${\Delta \hat{m}}_{n\bar{n}} = e_\nnBmm\, ({\Delta \hat{m}}_{n\bar{n}})_{\rm SK}$; 
the smaller the $e_\nnBmm$ the better is the experiment's sensitivity (with $e_\nnBmm\! =\! 1$ for SK).
From Eq.~(\ref{DmNNB.EQ}) we can write the \nnbar\ oscillation upper bound for our theory from a current or future experiment as 
\beq
10^{\kappa_\nnBmm} \equiv \frac{\seff^2}{e_\nnBmm} = 2.5\times 10^{-31} \left(\frac{13}{\QhME}\right) \left(\frac{M_\chi}{1~{\rm GeV}}\right)^5 \left[\frac{1}{g\, (M_\chi/\Lambda)^2}\right]^2  \ . 
\label{seffeNNBbnd.EQ}
\eeq
For a given experiment with a particular improvement factor $e_\nnBmm$, a nontrivial reach of the experiment exists in regions of
$M_\chi$, $g (M_\chi/\Lambda)^2$ parameter space that give $\seff \leq 1$ as computed from Eq.~(\ref{seffeNNBbnd.EQ}),
and the smaller the $\seff$ value so obtained, the better the reach.

We overlay in Fig.~\ref{YBVgMrMLSmp123AB.FIG} (right panel) contours of $\kappa_\nnBmm$
shown as the red slanted lines;
the same $\kappa_\nnBmm$ dependence also holds for the left and middle panels, which we do not show explicitly. 
To see a positive \nnbar\ oscillation signal in an experiment
for a parameter space point at the intersection of the black solid line ($\YB/\YBobs\!=\!1$ contour where the observed BAU is obtained)
with a particular $\kappa_\nnBmm$ red line, 
the sensitivity needed in the experiment is at least $e_\nnBmm = 10^{-\kappa_\nnBmm}$ so that $\seff=1$ is reached.
If the sensitivity is better than this, i.e. if the $e_\nnBmm$ value is smaller than $10^{-\kappa_\nnBmm}$, 
then the reach of that experiment is $\seff = \sqrt{e_\nnBmm}\, 10^{\kappa_\nnBmm/2}$. 
We see that for smaller $M_\chi$ values, a given $\kappa_\nnBmm$ curve intersects the $\YB/\YBobs\!=\!1$ contour at two points,
and a better reach for a given $e_\nnBmm$ is obtained for the upper branch of the BAU curve. 

The region to the left of the solid red line with $\kappa_\nnBmm\!=\! 0$ is already being nontrivially constrained by the SK limit,
obtained by setting $e_\nnBmm \!=\! 1$. 
For example, in BP-3B,
the SK bound is $\seff \lesssim 10^{-3/2}$ at $M_\chi \approx 10^{3.75}$~GeV
since the $\kappa_\nnBmm \!=\! -3$ contour just grazes the $\YB/\YBobs\!=\!1$ BAU contour at that $M_\chi$. 
Also, from the $\kappa_\nnBmm \!=\! 0$ contour for $\seff \!=\! 1$,
the SK bound is $M_\chi \!=\! 10^{4.75}$~GeV from the upper branch of the BAU contour. 

With improvements in the experimental sensitivity in the future,
for example with $e_\nnBmm \!=\! 10^{-3}$ (as for the expectation of the HIBEAM/NNBAR experiment mentioned above),
from the $\kappa_\nnBmm \!=\! -3$ contour, the corresponding reach improves to $\seff \!=\! 10^{-3}$ for $M_\chi \approx 10^{3.75}$~GeV; 
from the $\kappa_\nnBmm \!=\! 0$ contour, the reach improves to $\seff \!=\! 10^{-3/2}$ for $M_\chi \!\approx\! 10^{4.75}$~GeV
in the upper BAU branch,  
and,
for the $\kappa_\nnBmm \!=\! 3$ contour with $\seff \!=\! 1$, the reach is $M_\chi \!\approx\! 10^{5.5}$~GeV
in the same branch. 
The prospect in BP-3C is somewhat better;
for instance,
from the $\kappa_\nnBmm \!=\! 3$ contour with $\seff \!=\! 1$, the reach is $M_\chi \!\approx\! 10^{5.75}$~GeV. 

The prospect in BP-3A is not as good. 
The lowest mass where the observed BAU is obtained is around $10^6$~GeV
and the lowest $\kappa$ contour is $\kappa\!=\!6$,  
and so we need an experimental improvement of about six orders of magnitude over the SK bound in order to see a \nnbar\ oscillation signal
for the limiting case of $\seff\!=\! 1$.
To push to lower values of $\seff$, even bigger improvements in experimental sensitivity would be required. 

Thus, in our theory,
the lower end of the mass range where the observed BAU is obtained, namely $M_\chi \sim 10^4$--$10^6$~GeV,
is very interesting from the upcoming \nnbar\ oscillation experiments point of view.
One could also have signals of the $\Chi$ at a 100~TeV future hadron collider.
For a recent study along these lines, but with scalar operators, see for instance Ref~\cite{Davoudiasl:2015jja}.
From our discussion of the flavor aspects in Sec.~\ref{BGChiTh.SEC},
one could also look for effects of our new physics in flavor changing observables in present and future colliders such as
Belle-II, the Large Hadron Collider, or a future 100~TeV hadron collider.
In our theory the associated new colored quark-like fermions $\Up,\Dp$ can also be searched for directly at
future hadron colliders
(for some promising channels, see Refs.~\cite{Gopalakrishna:2011ef,Gopalakrishna:2013hua} and references therein). 
We relegate to future work the study of the prospects of these possibilities.

\section{Partial wash-out and the final $\nB$}
\label{nBFinWO.SEC}

In our theory, we have computed in Sec.~\ref{BEnum.SEC}, 
the baryon number yield that is generated by our baryogenesis mechanism and
has stabilized to its asymptotic value $\nB(t_e)\! \equiv\! \nBe$
at the time $t_e$ corresponding to $x_e\!=\!M_\chi/T_e$.
We have included in our work the effect of
decay and scattering processes, including $\Delta B\! =\! \pm 2$ scattering processes.

Independent of our baryon number generating mechanism,
if some other mechanism such as leptogenesis~\cite{Fukugita:1986hr} has, in addition, generated a nonzero lepton number density $n_L(t_e)\! \equiv\! \nLe$,   
we have $\nBmL(t_e)\!=\! \nBe-\nLe$ and $\nBpL(t_e)\!=\! \nBe+\nLe$.
At $t_e$, the baryon number generated in our mechanism is carried by the $\Up,\Dp$.
As discussed in Ref.~\cite{Gopalakrishna:2022hwk} and reviewed in Sec.~\ref{BGChiTh.SEC}, if the $\Up,\Dp$ mix with the SM quarks, 
they eventually decay and faithfully transfer this baryon number to SM quarks and leptons 
via $B$ and $L$ conserving interactions, 
and
the above $\nBmL$ and $\nBpL$ values persist even after their decay, now carried by the SM quarks and leptons. 

If other $B$ violating processes occur following this,
it may cause a partial wash-out of the generated $B$, leading to the survival fraction $w\! <\! 1$.
One such effect is the anomalous non-conservation of baryon number $B$~\cite{tHooft:1976rip}
due to electroweak sphalerons~\cite{Klinkhamer:1984di}, which violate $B+L$ but conserve $B-L$,
and is operative at the scale of the electroweak phase transition.
Ref.~\cite{Kuzmin:1985mm} claims that if sphalerons are unsuppressed, they drive $\nBpL\! \to\! 0$,
but in a detailed analysis, Ref.~\cite{Harvey:1990qw} argues that $B+L$ reduces but stabilizes at a nonzero value,
and if lepton number violation is also present, for example as Majorana masses of neutrinos,
the requirement that a nonzero baryon number survives places nontrivial constraints on the neutrino masses and the scale of $B-L$ violation.
We therefore allow for a nonzero terminal value of $B+L$ and take the surviving fraction to be not zero but $\alpha_w \nBpL$, with $0\!<\!\alpha_w\!<\!1$.
Thus, if the sphalerons are operative {\em after} time $t_e$, following its action, 
and after antiparticles have annihilated away leaving only particles, at time $t_s$,
we have $\nBpL(t_s)\! =\! \alpha_w \nBpL(t_e)$, and,
$\nBmL(t_s)\! =\! \nBmL(t_e)$ is unchanged since sphalerons conserve $B-L$.
Denoting $\nBs$ and $ \nLs$,
This implies
$\nB(t_s)\!\equiv\! \nBs\! =\! (1/2)[(1\!+\!\alpha_w) \nBe\! -\! (1\!-\!\alpha_w) \nLe]$ and
$n_L(t_s)\!\equiv\!\nLs\! =\!-(1/2)[(1\!-\!\alpha_w) \nBe\! -\! (1\!+\!\alpha_w) \nLe]$. 

Following such processes, after antiparticles have annihilated away,
we have a thermal bath of photons, protons, neutrons, electrons, neutrinos, and antineutrinos,
with number densities $n_\gamma$,~$n_p$,~$n_n$,~\allowbreak~$n_e$,~$n_\nu$,~$n_{\bar\nu}$ respectively,
where we allow for the possibility of a nonzero $n_{\bar\nu}$.
EM charge conservation ensures that $p^+$ and $e^-$ are in equal numbers, i.e. $n_p(t_s)\! =\! n_e(t_s)$,  
consistent with the EM charge neutral Universe we observe today.
We have
$n_p(t_s)+n_n(t_s) \!=\! \nBs\ $ and $n_e(t_s)+n_{\Delta\nu}(t_s)\!=\!\nLs$ where $n_{\Delta\nu}\! =\! n_\nu-n_{\bar\nu}$.
Defining $r_{np}\!\equiv\!n_n/n_p$, we have $\nps\!=\! \nBs/(1+\rnp)$ and $\nns\!=\! \nBs \rnp/(1+\rnp)$.
Thus, given the $\nBe$ and $\nLe$, and the $\alpha_w$ and $\rnp$ obtained from the sphaleron dynamics,
all the number densities are determined in terms of these. 
If we assume the isospin symmetric limit $n_p(t_s)\!=\! n_n(t_s)$, i.e. $\rnp\!=\!1$,
we obtain
$\nps\!=\!\nns\!=\!\nes\!=\!(1/4)[(1\!+\!\alpha_w) \nBe\! -\! (1\!-\!\alpha_w) \nLe]$ and
$\nDnus\!=\! -(3/4)[(1\!-\!\alpha_w/3)\nBe\! -\! (1\!+\!\alpha_w/3)\nLe]$.
In the limit of $\alpha_w\!\to\!0$ and $\nLe\!\to\!0$, we have a wash-out factor for $\nB$ of $w_{\rm sph}\! =\! 1/2$.

The total wash-out factor for $\nB$, namely $w_{\rm tot}$, is the product of the wash-out factors due to each such effect (if there are more), 
and the baryon number density following such baryon number violating processes
is $\nB(t_s)\! =\! w_{\rm tot} \nB(t_e)$, which is to be matched to the observed BAU.
For a given $w_{\rm tot}$, the parameter space point that leads to the observed BAU can be inferred from our results in
Sec.~\ref{BEnum.SEC}, in particular, Figs.~\ref{YBvGmSigSmp123AB.FIG},~\ref{YBVgMrMLSmp123AB.FIG},
by reading off the parameters corresponding not to the $\YB/\YBobs\!=\!1$ contour as we would have done ignoring wash-out,
but rather to the $1/w_{\rm tot}$ contour.

Our effective theory is to be UV-completed at around the $\Lambda$ scale,
and many completions are discussed in Ref.~\cite{Gopalakrishna:2022hwk} (see also Appendix~\ref{chiOutOfTEUV.SEC}), 
including, for instance, that the VV effective interaction is resolved by a vector state $\xi^c_\mu$.
Also, a simple example for the origin of the Majorana mass term from a baryon number conserving UV completion
is realized by adding a scalar field $\Phi_B$ with $B(\Phi_B)=-2$, 
whose nonzero vacuum expectation value (VEV) leads to the $\chi$ Majorana mass, breaking baryon number
(see Ref.~\cite{Gopalakrishna:2022hwk}, and Appendix~\ref{chiOutOfTEUV.SEC}). 
In the early Universe, at temperature $T \sim \Lambda$, the dynamics of the $\xi$, $\Phi_B$, etc. also becomes relevant, 
in that the decay or scattering of these states could also contribute a baryon asymmetry. 
However, we expect this to be washed out significantly due by the efficient scattering processes involving the $\Chi$
in TE.
These additional contributions, can only be computed accurately in specific UV completions; 
the spirit of our work here has been in computing the BAU in the effective theory common to all such UV completions.  

Following the epoch of BAU generation from the $\Chi$ decay and scattering we have focused on,
we connect to the standard cosmological scenario leading into big-bang neucleosynthesis (BBN),
and the standard post-BBN cosmology that is well constrained by observations.
This concludes our discussion of the baryon asymmetry generated in our theory, and a partial wash-out that may occur.

\section{Conclusions}
\label{Concl.SEC}

Our focus here has been the baryon asymmetry generated in the early Universe in our theory in Ref.~\cite{Gopalakrishna:2022hwk}
involving a Majorana fermion pair $\Chi$ coupled to quark-like fermions via a dimension-six four-fermion vector-vector (VV) interaction
of the form $(g/\Lambda^2) (\overline{\Dp^c} \, \gamma^\mu \Dp) \ ( \bar{\Chi}\gamma_\mu \Up)$.
Here we collectively denote as $Q\!=\!\{\Up,\Dp\}$ with baryon number $B(Q)\!=\!1/3$, 
and $\nB$ is then directly related to $\nQ - \nQB$, the difference in the $Q$ and the $Q^c$ number densities.

We set up the Boltzmann equations for the $\Chi$ yield $Y_\Chi(x)\!=\!\nX/s$ and the baryon number yield $\YB(x)\!=\!\nB/s$,
as a function of $x\!=\!M_\chi/T$ in the early Universe, 
where $\nX,\nB$ are the $\Chi$ and baryon number densities respectively, and $s$ is the entropy density.
We include in the collision terms of the Boltzmann equations 
the rates for the decay process $\Chi\to QQQ$,
the scattering process $\Chi Q^c\to QQ$,
and, $\Delta B\!=\! 2$ scattering process $QQQ \to Q^cQ^cQ^c$,
along with their conjugate and inverse processes and compute this asymmetry. 
We take the thermally averaged decay and scattering rates with their temperature dependence (i.e. $x$ dependence) from Ref.~\cite{Gopalakrishna:2023mul}
for the single-operator and multiple-operator contributions in benchmark points BP-A and BP-B respectively.
In addition, we define BP-C here to explore the possibility of the baryon asymmetry generated in $\Chi$ decay.
We thus have nine benchmark points BP-nX for n=\{1,2,3\} and X=\{A,B,C\} having the $x$-dependence in BP-A,\,B
as in Ref.~\cite{Gopalakrishna:2023mul}, 
while BP-C is with the same $x$ dependence as in BP-A but with a suppressed scattering rate to explore the possibility of
the decay contribution generating the baryon asymmetry. Such a situation may arise in other theories.
This allows us to explore the various aspects of the baryon asymmetry generation in a fairly general way
for a wide range of mass scale $M_\chi$.

Starting from an initially baryon symmetric Universe with $\YB(x_b)\!=\!0$,
we solve the Boltzmann equations numerically and 
find the regions of parameter space $M_\chi, g, \rMLtxt$ of our new physics theory 
for which $\YB(x_e)=\YBobs$, i.e. the observed BAU is obtained.
The baryon asymmetry is generated mainly in scattering processes in BP-A and BP-B,
and including these scattering rates in the Boltzmann equations is important.
Most other studies in the literature usually ignore scattering and include only decays. 
In BP-C we find that the asymmetry is generated mainly in decay processes.
We also present our results model independently by varying directly the $\DGm$, $\DSigv$ values, and show for which values the observed BAU is obtained.
Our most important finding here is that
the observed BAU is obtained for masses in a very wide range $M_\chi\!\sim\! (10^6,10^{16})$~GeV in BP-nA,
and $M_\chi\!\sim\! (10^4,10^{16})$~GeV in BP-nB and BP-nC.

%
%
We present the \nnbar\ oscillation rate over the viable parameter range, 
and show that recent experiments have already begun probing the viable lower range of $M_\chi$ values. 
Upcoming experiments will improve the mass reach, potentially uncovering such physics. 
%
%
High-$p_T$ and flavor probes at future colliders may also be sensitive to the lower range of $M_\chi$ values,
and we relegate to future work the identification of promising modes.


\appendix



\section{Review of Thermodynamics in the Early Universe}
\label{ThDUnivRev.SEC}

We collect here some well known thermodynamics expressions applicable in the early Universe that will be useful in this section. 
In thermal equilibrium (TE), the number density can be written as $n^{(eq)}(T) = e^{\mu/T} n^{(0)}(T)$ where $\mu$ is the chemical potential. 
The $n^{(0)}$ of a fermionic species,
the number density of the photon $n_\gamma$,
the entropy density $s$,
are given by~\cite{Kolb:1990vq,Dodelson:2003ft},
\bea
n_{NR}^{(0)}(T) &=& g \left(\frac{MT}{2\pi}\right)^{3/2} e^{-M/T} \  ({\rm for}\ T \ll M) \ ; \quad
n_{Rel}^{(0)}(T)  =  g \frac{3\zeta(3)}{4} \frac{1}{\pi^2} T^3 \ ({\rm for}\ T \gg M) \ , \label{neqOfTlt.EQ} \\
s(T) &=& \frac{2 \pi^2}{45}\, g_{*S}\, T^3 \ , \quad
n_\gamma = \frac{\zeta(3)}{\pi^2} g_\gamma T^3 \ , \quad
g_{*S}(T) \equiv \sum_{i={\rm Boson}}\!\!\!\! g_i \left(\frac{T_i}{T}\right)^3 + \frac{7}{8} \sum_{i={\rm Fermion}}\!\!\!\! g_i \left(\frac{T_i}{T}\right)^3 \ ,  \label{sngamOfT.EQ} \\
Y^{(0)}(T) &\equiv& \frac{n^{(0)}(T)}{s} \ ; \quad Y_{NR}^{(0)}(T) = \frac{45}{4\sqrt{2} \pi^{7/2}} \frac{g}{g_{*S}} \left(\frac{M}{T}\right)^{3/2} e^{-M/T} \ ; \quad
                     Y_{Rel}^{(0)}(T) = \frac{3 \zeta(3)}{4}\, \frac{45}{2 \pi^4} \frac{g}{g_{*S}} \ ,  \label{YeqOfTlt.EQ}
\eea
where
$Y \equiv n/s$, 
$g$ is the number of degrees of freedom (dof) of the species and
$g_\gamma = 2$ for the 2 photon polarizations,
$n_{NR}(T)$ is for temperatures below $M$ when the species is nonrelativistic,
the sum for $g_*$ is over relativistic species,
$\zeta(3) \approx 1.2$, 
and the factor of $3/4$ in $n_{Rel}$ and $Y_{Rel}$ is present if it is a fermionic species and is to be omitted if it is a bosonic species. 
From these we have $s = 0.44 g_{*S} T^3 = 1.8\, g_{*S}\, n_\gamma$.
Since the photon temperature changes due to reheating, $\eta_B$ is not constant,
and a better ratio to define~\cite{Kolb:1990vq} is 
$\YB \equiv (\nB - n_{\bar{B}})/s \approx \eta_B/7.04$ as it remains constant.

Starting from the Boltzmann limit of the distribution function,
we derive a generalization of Eq.~(\ref{neqOfTlt.EQ}) that is correct for any $T$ including for $T\sim M$,
and find
\beq
n^{(0)} = \frac{g}{2\pi^2} M^2 T\, K_2\left(\frac{M}{T}\right) \ ,
\label{neqOfT.EQ}
\eeq
where $K_2$ is the modified Bessel function of the second kind.
Using this, we readily obtain the generalization of Eq.~(\ref{YeqOfTlt.EQ}) as
\beq
Y^{(0)} = \frac{n^{(0)}}{s} = \frac{45}{4\pi^4}\, \frac{g}{g_{*S}}\, \left(\frac{M}{T}\right)^2 K_2\left(\frac{M}{T}\right) \ ,
\label{YeqOfT.EQ}
\eeq
valid for any $T$.
From the small and large argument limits of $K_2$ (see Ref.~\cite{Gopalakrishna:2023mul}),
we obtain Eq.~(\ref{YeqOfTlt.EQ}), up to the factor of $(3/4)\zeta(3) \approx 0.9$ in $Y_{Rel}^{(0)}(T)$,  
the difference being due to having taken the Boltzmann distribution limit.
Since the occupation numbers are small in the temperature range of interest to us,
we perform our calculations using the Boltzmann distribution limit. 

We take $g_*\! =\! g_{*S}\! =\! 106.75$ (due to SM dof, without $\nu_R$),
$g_\chi\!=\! 2$, $g_Q\! =\! 2\times 3$ (where $3$ is the color factor), $g_\Phi\! =\! 2$.
For the SM, taking all the SM fields to be massless, from the small argument limit of Eq.~(\ref{YeqOfT.EQ})
and taking $g\approx g_*\approx g_{*S}$, we find 
$\YSMz \approx (45/4\pi^4)\, (2) = 0.23$. 

In the radiation-dominated era, we can trade time $t$ to the temperature $T$ by using~\cite{Kolb:1990vq}
\beq
t \approx \frac{0.301}{\sqrt{g_*}} \frac{M_{Pl}}{T^2} \ ,
\label{tToTRD.EQ}
\eeq
where at the Planck time $t_{Pl} \equiv 1/M_{Pl}$ we assume that
the energy density $\rho = M_{Pl}^4$, and
$g_*$ counts the relativistic degrees of freedom,
and is given by a similar expression as $g_{*S}$ except we have $(T_i/T)^4$ on the rhs now instead of $(T_i/T)^3$.  
In this epoch, the Hubble expansion rate is given by
\beq
H(T) = 1.66\, \sqrt{g_*} \frac{T^2}{M_{Pl}} \ .
\label{HofT.EQ}
\eeq

From these expressions, the Boltzmann equations for $\nX,\nB$ that are of the form
\beq
\frac{d}{dt} n + 3 H n = (rhs) \ ,  
\eeq
can be written in terms of the {\em yield} $Y\!=\! n/s$, for each of $Y_\chi,\YB$, and $x\!=\!M_\chi/T$ variables as
\beq
\frac{d}{dt} Y = \frac{1}{s} (rhs) \ , \quad {\rm or, } \quad \frac{d}{dx} Y = x \frac{1}{H(M_\chi)} \frac{1}{s} (rhs) \ ,  
\eeq
where $H(M_\chi) = 1.66 \sqrt{g_*} M_\chi^2/M_{Pl}$.

\section{Out of thermal equilibrium $\Chi$ and UV completion}
\label{chiOutOfTEUV.SEC}

If the $\Chi$ is not in TE, say at $x_b$, what initial condition $n_\chi(x_b)$ is to be taken
for solving the BE becomes a question.
Of all the BPs considered, this scenario is encountered only in BP-1B (cf. Sec.~\ref{BEnum.SEC}). 
This signals that the generation of baryon asymmetry becomes dependent on the UV completion details.
We consider an example UV completion, and abstract from this a parametrization
that is applicable to other UV completions also. 

Consider an example UV completion (cf. Ref~\cite{Gopalakrishna:2022hwk}), 
in which $B$ is conserved, but only spontaneously broken.
We have
${\cal L}_{\rm int} \supset -(\tilde{y}/\sqrt{2})\, \Phi_B \overline{\Chi^c} \Chi - (y'/\sqrt{2})\, \Phi' \overline{\Chi} \Chi + {\rm h.c.}$
where $\Phi_B$ is a complex scalar with $B(\Phi_B) = -2$ and mass $M_\Phi$, and $\Phi'$ another scalar with $B(\Phi') = 0$ and mass $M_{\Phi'}$.
The potential is such that they acquire VEVs $\left<\Phi_B\right> = v_\Phi/\sqrt{2}$ leading to the $\Chi$ Majorana mass $\tilde{M} = \tilde{y} v_\Phi$, 
and $\left<\Phi'\right> = v'_\Phi/\sqrt{2}$ leading to the $\Chi$ Dirac mass $M_\chi = y' v'_\Phi$.
We are in the pseudo-Dirac limit if $M_\chi \approx \tilde{M}$.
The VV effective operators we have been working with in this study are generated by integrating out other states at the cutoff scale $\Lambda$.

The coupling of the $\Phi_B$, $\Phi'$ with the SM could be, for example, via the SM Higgs $H$ via 
${\cal L}_{\rm int} \supset - (\kappa \, \Phi_B^* \Phi_B + \kappa' \, {\Phi'}^* \Phi') H^\dagger H$.
In contrast, there are no renormalizable operators directly coupling the $\Chi$ to the SM fields that conserve baryon number and the SM gauge symmetries. 
It is perhaps worth clarifying that the one operator which could have done this, namely, $\bar\Chi L\cdot H$, is also disallowed since $\Chi$ carries nonzero $B$ number. 
If $\kappa,\kappa'$ are sizable, the $\Phi_B,\Phi'$ are kept in TE with the SM,
and if $\tilde{y},y' \ll 1$ the $\Chi$ may not be in TE.
In this case a significant $\Chi$ number density builds-up only after the $\Phi \to \Chi \Chi$ decay epoch,
provided $M_\Phi \gtrsim 2 M_\Chi$.
Let us assume that this situation holds.
Here, we denote by $\Phi$ either of the $\{\Phi_B,\Phi'\}$ as it is not important to keep the distinction below. 

Following $\Phi \to \Chi \Chi$ decays, 
we can write from energy conservation, $M_\Phi n_\Phi \approx M_\chi \nX$, assuming that the $\Phi,\Chi$ are not too relativistic. 
Thus $\nX \approx (M_\Phi/M_\Chi) n_\Phi^{(eq)}(T)$ owing to our assumption that $\Phi$ was in TE before it decayed.
This implies a $\Chi$ overdensity at $x=x_b\equiv M/\Lambda$ of
\beq
\delta_\Chi(x_b) \equiv \nX(x_b)/\nXeq(x_b) = Y_\Chi(x_b)/Y_\Chi^{(eq)}(x_b) = (M_\Phi/M_\Chi) Y_\Phi^{(eq)}(x_b)/Y_\Chi^{(eq)}(x_b) \ .
\label{deltaChiPhi.EQ}
\eeq
If $\Chi$ is decoupled, the relaxation time for this $\Chi$ overdensity could be significantly long.
Since the exact value of $\delta_\Chi$ is dependent on the UV completion details,
we present our results for various values of $\delta_\Chi$ for BP-1B.

\section{The decay rate and scattering cross sections in the benchmark points}
\label{GmSigvxDep.SEC}

Here we collect the functional forms of the thermally averaged decay and scattering rates and cross sections 
$\GmTA,\DGmTA,\SigvTA,\DSigvTA,\GmPTA$ as a function of $x=M_\chi/T$ that we have used in our numerical work.
From the results of Ref.~\cite{Gopalakrishna:2023mul}, we can extract powers of $M_\chi$ and $\Lambda$,
and define dimensionless rates and cross sections as
\bea
\GmzTA &=& |g|^2 \frac{M_\chi^5}{\Lambda^4}\,\GmzTAh \ , \quad
\DGmzoTAh = {\rm Im}(g^4)\,\left\{ \frac{M_\chi^9}{\Lambda^8} \ , \frac{M_\chi^7}{\Lambda^6} \right\} \DGmzoTAhh \ , \\
\SigvzTA &=& |g|^2 \frac{M_\chi^2}{\Lambda^4}\,\SigvzTAh \ , \quad
\DSigvzoTAh = {\rm Im}(g^4)\,\left\{ \frac{M_\chi^6}{\Lambda^8}, \frac{M_\chi^4}{\Lambda^6} \right\} \DSigvzoTAhh \ , \\
\GmzPTA &=& |g|^4\, \frac{M_\chi^9}{\Lambda^8}\,\GmzPTAh \ ,
\eea
where
we have scaled the rates with appropriate powers of $g$, 
and shown as $\{.~, .\}$ the single operator (BP-A) and multiple operator (BP-B) contributions respectively.
We note the extra suppression of $M_\chi^2/\Lambda^2$ in $\DGmzoTAh,\DSigvzoTAhh$ for BP-A relative to BP-B. 

The following functions mimic adequately well the magnitude and $x$ dependence of the rates and cross sections computed in Ref.~\cite{Gopalakrishna:2023mul}
for the single and multiple operator benchmark points BP-A and BP-B respectively.
\bea
\GmzTAh(x) &=& 1.7\times 10^{-6}\, \frac{K_1(\hat{M} x)}{K_2(\hat{M} x)} \ , \quad 
\DGmzoTAhh(x) = -1.4\times 10^{-9}\, \frac{K_1(\hat{M} x)}{K_2(\hat{M} x)} \ 
\ {\rm (for\ BP\!\!-\!\!A)} \ , \label{GmTABPAB.EQ} \\
\GmzTAh(x) &=& 2\times 10^{-5}\, \frac{K_1(\hat{M} x)}{K_2(\hat{M} x)} \ , \quad 
\DGmzoTAhh(x) = -6\times 10^{-11}\, \frac{K_1(\hat{M} x)}{K_2(\hat{M} x)} \ 
\ {\rm (for\ BP\!\!-\!\!B)} \ . \nonumber
\eea
For the thermally averaged scattering cross section, combining SC-1 and SC-2, 
we take as a fit function a Gaussian core function with a power-law tail given by
\bea
\SigvzTAh(x) \!\! &=&\!\!  12.7\, \left\{ \begin{array}{cc} 
  0.1+0.9\, e^{-(x-0.5)^2/0.25^2} & {x \leq 0.5} \\
  0.001/x+(1-0.001/x)\, e^{-(x-0.5)^2/0.5^2} & {x > 0.5}
                \end{array} \right. {\rm (for\ BP\!\!-\!\!A)}  \ , \label{SigTABPAB.EQ} \\ 
\DSigvzoTAhh(x) \!\! &=&\!\! - 0.04\, \left\{ \begin{array}{cc} 
   e^{-(x-0.5)^2/0.25^2} & {x \leq 0.5} \\
   0.1/x+(1-0.1/x)\, e^{-(x-0.5)^2/0.75^2} & {x > 0.5}
                \end{array} \right. {\rm (for\ BP\!\!-\!\!A)} \ , \nonumber \\
\SigvzTAh(x) \!\! &=&\!\! 3.5\,\left\{ \begin{array}{cc} 
  e^{-(x-0.5)^2/0.15^2} & {x \leq 0.5} \\
  0.01/x+(1-0.01/x)\, e^{-(x-0.5)^2/0.5^2} & {x > 0.5}
                \end{array} \right. {\rm (for\ BP\!\!-\!\!B)}  \ , \nonumber \\ 
\DSigvzoTAhh(x) \!\! &=&\!\! - 0.3\, \left\{ \begin{array}{cc} 
   e^{-(x-0.5)^2/0.18^2} & {x \leq 0.5} \\
   0.1/x+(1-0.1/x)\, e^{-(x-0.5)^2/0.5^2} & {x > 0.5}
                \end{array} \right. {\rm (for\ BP\!\!-\!\!B)} . \nonumber  
\eea
For the $UDD\to U^cD^cD^c$ process we take the thermally averaged scattering rate
\bea
\GmzPTAh = \left\{ \begin{array}{cc}  
  0.4\,\left[0.25\, K_1(\hat{M} x)/K_2(\hat{M} x) + e^{-(x-0.25)^2/0.15^2}\right] & {\rm (for\ BP\!\!-\!\!A)} \\ 
  6\,\left[K_1(\hat{M} x)/K_2(\hat{M} x) + 0.1\, e^{-(x-0.25)^2/0.15^2}\right] & {\rm (for\ BP\!\!-\!\!B)}
                \end{array} \right.   \ . \label{GmPTABPAB.EQ} 
\eea



\end{document}